\documentclass[a4paper,12pt]{article}

\usepackage{amsmath}
\usepackage{amssymb}
\usepackage[dvips]{graphicx}  
\begin{document}

\begin{titlepage}
\begin{flushright}
IFUP--TH 2004/20 \\
\end{flushright}
~

\vskip .8truecm

\begin{center}
\Large\bf
Standard and geometric approaches \\
to quantum Liouville theory on the  pseudosphere
\footnote{This work is  supported in part
  by M.I.U.R.}
\end{center}

\vskip 2truecm
\begin{center}
{Pietro Menotti} \\ 
{\small\it Dipartimento di Fisica dell'Universit{\`a}, Pisa 56100, 
Italy and}\\
{\small\it INFN, Sezione di Pisa}\\
{\small\it e-mail: menotti@df.unipi.it}\\
\end{center}
\vskip .8truecm
\begin{center}
{Erik Tonni} \\  
{\small\it Scuola Normale Superiore, Pisa 56100, Italy and}\\
{\small\it INFN, Sezione di Pisa}\\
{\small\it e-mail: e.tonni@sns.it}\\
\end{center}

\vskip 2truecm

\begin{abstract}

We compare the standard and geometric approaches to quantum 
Liouville theory on the pseudosphere by performing perturbative 
calculations of the one and two point functions up to the third 
order in the coupling constant.
The choice of the Hadamard regularization within the
geometric approach leads to a discrepancy with the standard approach.
On the other hand, we find complete agreement between the results
of the standard approach and the bootstrap conjectures 
for the one point function and the auxiliary two point function.
 
\end{abstract}

\vskip 1truecm

\end{titlepage}

\section*{Introduction}

Classical Liouville theory is well understood
even if it can be solved explicitly only in special cases.
Quantum Liouville theory was first developed in the hamiltonian
framework \cite{CT, DFJ}. Later,
functional techniques were  
applied to the euclidean version of the theory.

Within the functional formulation there exist the so called standard
approach \cite{ZZ:Pseudosphere,FZZ,ZZ:Sphere,Dorn Otto,Teschner,GoulianLi} 
and the geometric approach \cite{Takhtajan:Topics}. In the first
case one introduces the vertex function by adding to the Liouville
action external currents as is usually done in quantum field
theory. Within the geometric approach one starts from the regularized
classical action in presence of boundary terms that represent the
sources and then considers the fluctuations of the field around the
classical background.

In this paper we shall compare the standard and the geometric
formulations of quantum Liouville theory on the sphere and the
pseudosphere with particular attention to the pseudosphere
\cite{ZZ:Pseudosphere},
where the perturbative results are directly compared.
Both formulations need to be regularized and such a regularization
process is crucial, since the results depend non trivially on the
adopted regularization procedure.
                                                                               
In the geometric approach the action is defined through a limit
process and such a structure is not so easy to use in explicit
calculations. However it is possible, by introducing a background field
and a source field, to rewrite the action in a form such that
no limit procedure appears.
This structure is not as elegant as the original one,
but from it one can read directly the transformation properties of the
off shell action and consequently also of the correlation functions.
This form can be used
as a starting point for perturbative calculations
of the correlation functions of vertex operators on the pseudosphere.

As first shown in \cite{Takhtajan:Topics},
the appealing properties of the geometric action
is to transform under conformal transformations as
the vertex correlation functions of the quantum theory, generating
quantum conformal 
dimensions  $\Delta_\alpha =\alpha\,(Q-\alpha)$, which are those found in
the hamiltonian approach, provided the central charges are properly
identified.
                                                                               
On the sphere, due to the bounds imposed by the Picard inequalities,
the situation is more complex and in this paper we
shall perform perturbative computations only on the pseudosphere.
                                                                               
With regard to the one point function the main outcome will be the
following: the results of the two formulations agree up to order $b^4$
included for the values of the first cumulant $G_1$, provided one
properly identifies the coupling constant $b_\textrm{g}$ of the
geometric approach as a function of the coupling constant $b$ of the
standard approach and the same for the cosmological constants. In
order to match the value of the second cumulant $G_2$, one has to
introduce by hand a coupling constant dependence in 
the source subtraction term of the geometric action. 
Such a $b_\textrm{g}$ dependent subtraction influences
only the second cumulant and has no effects on the other cumulants
(\,$G_1$ and $G_n$ with $n\geqslant 3$\,). A significant test of the
equivalence of the two perturbative expansions can be achieved by
computing the third cumulant $G_3$. This is easily done to first order,
yet some improvement in the computational technique is needed both in
the standard and geometric approach, in order to get the third
order coefficient.  The explicit computation of the third cumulant to
order $b^3$ within the standard approach has been given in
\cite{tetrahedron} and it disagrees with the result of the geometric
approach computed in Section \ref{One point function} of the present paper.
                                                                              
From the general field theoretical point of view we find that within
the geometric approach the asymptotic value of the the vacuum
expectation value of the Liouville field reproduces the classical
background value.  This does not happen within the standard approach,
where the two asymptotic behaviors agree only qualitatively.  On the
other hand, within the standard approach the field $e^{2b\phi}$,
which appears in the cosmological term, transforms like a $(\,1,1\,)$
primary field, while within the geometric approach 
the analogous field $e^{2b_\textrm{g}\phi}$
has not such a transformation property because its quantum
conformal dimensions are  $(\,1-b_g^2,1-b_g^2\,)$. 
The different characters of the
operators appearing in the cosmological terms of the two approaches
has been already noticed by Takhtajan \cite{Takhtajan:Equivalence}.  Here
we find a difference respectively at the second and third order in the
perturbative expansion of the second and third cumulant and these
differences cannot be matched consistently by a redefinition of the
coupling constants.  

The obtained results can be compared to the
perturbative expansion of the formula conjectured by Zamolodchikov and
Zamolodchikov (\,ZZ\,) for the one
point function on the pseudosphere \cite{ZZ:Pseudosphere}.  Complete
agreement has been found with the perturbative computation within the
standard approach up to the third order \cite{tetrahedron}.  To gain
deeper insight, we shall move further through
the perturbative computation of the two
point function on the pseudosphere. This allows not only a better 
comparison between
the two approaches but also to compare the results with another
conjecture \cite{ZZ:Pseudosphere, FZZ, Teschner}, i.e. the two point
function when one vertex is the degenerate field $e^{-b\phi}$.  The
interest of this computation is that, by taking a proper ratio of two
and one point functions, we can extract quantities  not
depending on the possible ambiguities in the subtraction terms of the
geometric approach. Once more we find that the two point functions are
different within the two approaches and that the perturbative results
of the standard approach agree with the perturbative expansion of the
formula conjectured through the bootstrap method.  

A deeper inspection shows that
the origin of the differences between the results obtained within the
two approaches lies in the different regulators employed and not
in the way used to introduce the sources.  Indeed, it can be shown that
adopting the ZZ regulator both the  approaches produce the same results,
identifying exactly the couplings and the cosmological constants. 
The discrepancy in the structure of the unperturbed dimensions is matched
by the different origin of the zero order in the second cumulant of the
one point function. For all the other orders
there is a one to one correspondence between the contributions.\\

\section{Geometric action on the pseudosphere}
\label{geometric action on pseudosphere}

 The pseudosphere can be represented on the upper half plane
 or on the unit disk $\Delta$. We shall use mostly the
 disk representation.\\
 At the quantum level the geometry is encoded by the boundary
 condition at $\infty$, i.e. on the unit circle.\\
 Within the geometric approach, we assume that the Liouville 
 field $\phi$ behaves like
\begin{eqnarray}
 \label{phi pseudosphere at infty}
\phi \hspace{.2cm} \simeq  \hspace{.2cm}
-\,\frac{Q}{2}\,\log\,(\,1-z\bar{z}\,)^2+\, O(1) 
& &
 \hspace{1cm} |z| \rightarrow \,1 \\ 
 \label{phi pseudosphere at sources}
\phi \hspace{.2cm} \simeq \hspace{.1cm}  
-\,\alpha_n \log\,|\,z-z_n\,|^2\,+\, O(1) 
& & 
 \hspace{1cm} \phantom{|}z \phantom{|} \rightarrow \,z_n~.
\end{eqnarray}
The $N$ point vertex functions are  defined as follows
\cite{Takhtajan:Topics, Takhtajan:Equivalence}
\begin{equation}
 \label{geometric N-point function}
\left\langle \, V_{\alpha_1}(z_1)\dots V_{\alpha_N}(z_N)  \,\right\rangle\,=\,
\left\langle\,e^{2\alpha_1 \phi(z_1)}
\dots e^{2\alpha_N \phi(z_N)}\,\right\rangle 
 \,=\,\frac{\displaystyle \int_{\mathcal{C}(\Delta)}
\mathcal{D}\,[\, \phi \,]\,\, 
e^{-S_{\Delta,\,N} 
\left[ \,\phi \,\right] }}{\displaystyle \int_{\mathcal{C}(\Delta)}
\mathcal{D}\,[\, \phi \,]\,\, 
e^{-S_{\Delta,\,0} 
\left[ \,\phi \,\right] }} 
\end{equation}
where $S_{\Delta,\,N} \left[ \,\phi \,\right]$ 
is the geometric action on the pseudosphere with $N$ sources 
\begin{eqnarray} 
  \label{geometric action Pseudosphere}
\hspace{-1cm}
S_{\Delta,\,N}[ \,\phi\,] \rule{0pt}{.9cm}& = &
 \lim_{\begin{array}{l}
\vspace{-.9cm}~\\
\hspace{.02cm} \vspace{-.4cm} \scriptscriptstyle \varepsilon \rightarrow 0\\
\hspace{.1cm} \scriptscriptstyle \!\!r \rightarrow 1\end{array}}\,
\Bigg\{\;\int_{\Delta_{r,\varepsilon}} \left[
\,\frac{1}{\pi}\, \partial_z \phi \,\partial_{\bar{z}}\phi+\mu_\textrm{g}
e^{2b_\textrm{g}\phi}\,\right]\, d ^2 z \, \nonumber \\
  &  & \hspace{1.5cm}
-\,\frac{Q}{2\pi i} \oint_{\partial\Delta_r} \phi
\left( \, \frac{\bar{z}}{1-z\bar{z}} \,d z- 
       \,\frac{z}{1-z\bar{z}} \,d \bar{z} \right)+ f(r,b_{\textrm{g}})
       \nonumber \\  
  &  & \hspace{1.5cm} \rule{0pt}{1cm}
-\,\frac{1}{2\pi i}\sum_{n=1}^N \alpha_n\oint_{\partial\gamma_n} \phi
\left( \, \frac{d z}{z-z_n}- \frac{d \bar{z}}{\bar{z}-\bar{z}_n}\,
\right)- \sum_{n=1}^N \alpha_n^2 \log \varepsilon_n^2 
\hspace{.2cm}\Bigg\}~.
\end{eqnarray}
\rule{0pt}{.2cm}\\
The points $z_n \in \Delta$, for $n=1 \dots N$, 
are the positions of the sources;
the domains of integration are $\Delta_r=\{ |z| \leqslant r \}$, 
$\gamma_n=\{ |z-z_n| \leqslant \epsilon_n \}$ and $\Delta_{r,\varepsilon}=
\Delta_r \backslash \bigcup_n \gamma_n$, 
while $f(r,b_{\textrm{g}})$ is a numerical subtraction term.
$S_{\Delta,0}\left[ \,\phi \,\right]$ is the action 
(\ref{geometric action Pseudosphere}) in absence of
sources, which is formally equal to the action of the standard
approach.\\
The equation of motion is the Liouville equation
in presence of sources 
\begin{equation}
 \label{Liouville eq. sources pseudosphere}
\partial_z \partial_{\bar{z}}\phi = \pi b_\textrm{g} \,\mu_\textrm{g} 
e^{2b_\textrm{g}\phi}
-\pi \sum_{n=1}^N \alpha_n \delta^2(\,z-z_n\,)~.
\end{equation}
We decompose $\phi$ as the sum of two classical fields 
(\,$\phi_B$ and $\phi_0$\,) and a quantum field $\phi_M$
\begin{equation}
 \label{phi pseudosphere decomposition phi0}
\phi= \phi_M + \phi_0 + \phi_B
\end{equation}
with the background field $\phi_B$ 
having the asymptotics (\ref{phi pseudosphere at infty}), i.e.
\begin{eqnarray}
 \label{phiB pseudosphere at infty}
\phi_B  \simeq -\,\frac{Q}{2}\,\log\,(\,1-z\bar{z}\,)^2 +c_{B,\Delta}+ 
o\left(\,1-|z|\,\right) 
\qquad \hspace{1cm}  |z| \rightarrow 1
\end{eqnarray}
where $c_{B,\Delta}$ is a constant.\\
One could choose for the source field $\phi_0$ a solution of the
equation
\begin{equation}
 \label{eq. phi0 pseudosphere}
\partial_z \partial_{\bar{z}}\phi_0 = -\,\pi \sum_{n=1}^N
 \alpha_n\, \delta^2(\,z-z_n\,)
\end{equation}
similarly to what has been done on the sphere 
(\,see \cite{CMS:Liouville} and Appendix\,).\\
Requiring that $\phi_0$ vanishes on the unit circle, one gets
\begin{equation}
 \label{phi0 pseudosphere}
\phi_0= -\sum_{n=1}^N \alpha_n \,
\log \,\left|\,\frac{z-z_n}{1-z\bar{z}_n}\,\right|^2.
\end{equation}
However, such $\phi_0$ vanishes too slowly for $|z| \rightarrow 1$,
giving rise to ill defined integrals in the perturbative calculation.\\
Thus, instead of $\phi_0$ given by 
(\ref{phi0 pseudosphere}), we shall choose $g_0$ satisfying
\begin{equation}
 \label{eq.for g_0} 
 \,\left(\,  \partial_z \partial_{\bar{z}}\,-
 2 \, \frac{1}{ \, (\,1-z\bar{z}\,)^2}\, \right)\,g_0 = 
 \,-\,\pi\,\sum_{n=1}^N \alpha_n \,\delta^2 \,(\,z-z_n \,)~.
\end{equation}
The solution of this equation is
\begin{equation}
 \label{g_0}
g_0(z;z_1 \dots z_N)= 2\,\sum_{n=1}^N \alpha_n \, g(z,z_n)
\end{equation}
being $g(z,z')$ the propagator \cite{DFJ, ZZ:Pseudosphere}
\begin{equation}
 \label{g(z,z_n)}
g(z,z') =  \,  -\, \frac{1}{2}\, 
\left(\,\frac{1+\,\eta}{1-\,\eta}\, \log \,\eta \, +2\,\right)
\end{equation}
and $\eta(z,z')$ the $SU(1,1)$ invariant
\begin{equation}
 \label{eta(z,z_n)}
\eta (z,z')= \left|\,\frac{z-z'}{1-z\bar{z}'}\,\right|^2
\end{equation}
which is related to the geodesic distance between $z$ and $z'$.\\
The source field $g_0$ converges to zero like
$O\big(\,(1-z\bar{z})^2\,\big)$ when $|z| \rightarrow 1$,
which makes the perturbative integrals convergent at infinity.\\
A procedure similar to the one employed for the sphere 
(\,see \cite{CMS:Liouville} and Appendix\,) gives  the
geometric action on the pseudosphere with a generic background field
$\phi_B$ that satisfies the boundary conditions 
(\ref{phiB pseudosphere at infty}). The result is
\begin{eqnarray} 
  \label{geometric action pseudosphere phiM}
S_{\Delta,\,N}[ \,\phi\,] & = & S_{\Delta,B}[ \,\phi_B\,]+ 
S_{\Delta,N,M}[ \,\phi_M,\phi_B\,]
  + \sum_n^N \alpha_n \sum_{m \neq n}^N \alpha_m \left[\,
  \frac{1+\,\eta_{n,m}}{1-\,\eta_{n,m}}\, \log \,\eta_{n,m} \, +2\,\right]
  \nonumber \\ 
 & & -\sum_n^N \alpha_n^2 \,
\left[\, \log\,(\,1-z_n\bar{z_n}\,)^2-2\, \right]
  -2\,\sum_n^N \alpha_n \,\phi_B(z_n) 
\end{eqnarray}
where $\eta_{n,m}=\eta (z_n,z_m)$ and 
$S_{\Delta,B}[ \,\phi_B\,]$ is the background action
\begin{eqnarray} 
 \label{standard action Pseudosphere phiB}
\hspace{-2.2cm}
S_{\Delta,B}[ \,\phi_B\,] & = & \lim_{r \rightarrow 1}
\Bigg\{\;\int_{\Delta_r} \left[
\,\frac{1}{\pi} \,\partial_z \phi_B \,\partial_{\bar{z}}\phi_B+\mu_\textrm{g}
e^{2b_\textrm{g}\phi_B}\,\right]\, d ^2 z  \nonumber \\
  &  & \hspace{1.6cm} -\frac{Q}{2\pi i} \oint_{\partial\Delta_r} \phi_B
\left( \, \frac{\bar{z}}{1-z\bar{z}} \,d z- 
       \,\frac{z}{1-z\bar{z}} \,d \bar{z} \right)+
       f(r,b_{\textrm{g}}) \;\Bigg\}  
\end{eqnarray}
while $S_{\Delta,N,M}[ \,\phi_M,\phi_B\,]$ is the action for the
quantum field $\phi_M$
\begin{eqnarray}
 \label{action phiM pseudosphere}
\hspace{-.55cm}
S_{\Delta,N,M}[ \,\phi_M,\phi_B\,]  \hspace{-.1cm}
&=& \hspace{-.2cm}
\int_{\Delta}  \left[\,\,\frac{1}{\pi}\,
  \partial_z  \phi_M \,\partial_{\bar{z}}\phi_M +
  \mu_\textrm{g} e^{2b_\textrm{g} \phi_B}
\Big(e^{2b_\textrm{g}(\,\phi_M+g_0\,)}-1 \Big)\right.
\nonumber \\   
& & \hspace{.8cm}
\left. -\,\frac{2}{\pi}\,(\,\phi_M+g_0\,)\,\partial_z \partial_{
\bar{z}} \phi_B -\,\mu_\textrm{g} e^{2b_\textrm{g} \phi_{B}^{cl}}\,
2b_\textrm{g}^2\,g_0\,
(\,g_0+2\phi_M\,)\,\,\right]\,d ^2z~.  \nonumber \\
& &  
\end{eqnarray}
The classical background field $\phi_B^{cl}$ is given by
\begin{equation}
 \label{phiBcl pseudosphere}
\phi_B^{cl} (z)= -\,\frac{1}{2b_\textrm{g}}\,\log\,
\left[ \, \pi b_\textrm{g}^2 \mu_\textrm{g} \,
  (\,1-z\bar{z} \,)^2 \, \right] 
\end{equation}
and it solves the Liouville equation
\begin{equation}
\partial_z \partial_{\bar{z}}\phi_B^{cl}
=\pi b_\textrm{g}\mu_\textrm{g}e^{2 b_\textrm{g} \phi_B^{cl}}
\end{equation}
with boundary conditions 
(\ref{phiB pseudosphere at infty}) if $Q=1/b_\textrm{g}$. The field
$\phi_B^{cl}$ enters into $S_{\Delta,N,M}[ \,\phi_M,\phi_B\,]$ 
because of the introduction of the singular field $g_0(z;z_1\dots z_N)$
through the equation (\ref{eq.for g_0}).\\
Under $SU(1,1)$ transformations
\begin{equation}
 \label{SL(1,1) transformation}
z \hspace{.2cm}\longrightarrow \hspace{.2cm} w= \frac{az+b}{\bar{b}z+\bar{a}} 
\hspace{1cm} \qquad 
|a|^2-|b|^2=1
\end{equation}
the background field transforms as follows
\begin{eqnarray}
 \label{phiB prime pseudosphere}
\phi_B(z) & \rightarrow & \phi_B ' (w)= \phi_B(z)-
\frac{Q}{2}\log \left| \frac{d w}{d z}\right|^{2}
\end{eqnarray}
while $\phi_0$ and $\phi_M$ are scalars. As a result,
the background action (\ref{standard action Pseudosphere phiB}) 
with $Q=1/b_\textrm{g}$ is $SU(1,1)$ invariant. 
The action $S_{\Delta,N,M}[ \,\phi_M,\phi_B\,]$ is invariant as well
and the terms that transform in (\ref{geometric action pseudosphere phiM}) are 
\begin{equation}
 \label{conformal dimensions terms}
 -\sum_n^N \alpha_n^2 \,\left[\; \log(\,1-z_n\bar{z}_n\,)^2 \; \right]
  -2\,\sum_n^N \alpha_n \,\phi_B(z_n)~. 
\end{equation}
Therefore the transformation law for the geometric action under $SU(1,1)$ is
\begin{equation}
 \label{action pseudosphere transformation}
S'_{\Delta, \,N}[ \,\phi'\,] = S_{\Delta, \,N}[\,\phi\,] + 
\sum_{n=1}^N \alpha_n\,(\,Q-\alpha_n\,)\, \log \left|\, \frac{d w}{d z} \, 
\right|_{z=z_n}^2
\end{equation}
where $Q=1/b_\textrm{g}$.\\
It is important to observe that such a trasformation property
for the action holds off shell, i.e. also when $\phi$ is not a solution
of the equation of motion.\\ 
With respect to the integration measure, we could choose the one
induced by the distance
\begin{equation} 
 \label{dphi distance e^2bphi}
(\,\delta \phi \,,\delta \phi \, )=
  \int_{\Delta}|\,\delta \phi \,|^2 e^{2b_\textrm{g}\phi}\, d^2 z
\end{equation}
or by the distance
\begin{equation} 
 \label{dphi distance e^2bphiB} 
(\,\delta \phi \,,\delta \phi \, )=
  \int_{\Delta}|\,\delta \phi \,|^2 e^{2b_\textrm{g}\phi_B}\, d^2 z~.
\end{equation}
Both are invariant under the group $SU(1,1)$, when $\phi_B$ transforms
like (\ref{phiB prime pseudosphere}) with $Q=1/b_\textrm{g}$,
but the measure induced by (\ref{dphi distance e^2bphi}) 
is not invariant under translation of the Liouville field $\phi$,
while the other one is.\\
The measure induced by (\ref{dphi distance e^2bphi}) differs from the
one induced by (\ref{dphi distance e^2bphiB}) by
ultralocal terms. Such a difference should not be relevant in 
perturbative calculation \cite{'t Hooft}. We will work with 
(\ref{dphi distance e^2bphiB}), which gives rise to an integration measure
that is invariant under translations in the field $\phi$.\\
Thus, we have that
\begin{equation}
\displaystyle \int_{\mathcal{C}(\Delta)}
\mathcal{D}\,[\, \phi \,]\,\, e^{-S_{\Delta,N,M}\left[
\,\phi_M,\,\phi_B\,\right]}  
\hspace{.2cm}=
\displaystyle \int_{\mathcal{C}(\Delta)}
\mathcal{D}\,[\, \phi_M \,]\,\, e^{-S_{\Delta,N,M}\left[
\,\phi_M,\,\phi_B \,\right]} 
\end{equation}
and it is invariant under $SU(1,1)$, for every $N$.\\
From the transformation law (\ref{action pseudosphere transformation}), 
one derives the quantum conformal dimensions 
of the Liouville vertex operators $e^{2\alpha\phi}$
\begin{equation}
 \label{conformal dimensions pseudosphere}
\Delta_{\alpha} = \alpha\,\left(\,Q-\alpha\,\right)=
\alpha\,\left(\,\frac{1}{b_\textrm{g}}-\alpha\,\right)~.
\end{equation}
\noindent On the classical background $\phi_B=\phi_B^{cl}$ given in
(\ref{phiBcl pseudosphere}) $S_{\Delta,N,M}[ \,\phi_M,\phi_B\,]$
can be written as follows
\begin{eqnarray} 
 \label{action phiM N=1}
\rule{0pt}{.7cm}
S_{\Delta,N,M}[ \,\phi_M,\phi_B^{cl}\,] & = &  \nonumber \\
& &
\rule{0pt}{1cm}\hspace{-4cm}= \hspace{.2cm}
\int_{\Delta}  \left[\,\frac{1}{\pi}
  \partial_z  \phi_M \partial_{\bar{z}}\phi_M +
 \mu_\textrm{g}e^{2b_\textrm{g} \phi_B^{cl}}
 \Big(e^{2b_\textrm{g}(\,\phi_M+g_0\,)}-1 -2b_\textrm{g}(\,\phi_M+g_0\,)\Big)
  \right. \nonumber \\
& &\hspace{4.4cm}
\left. \phantom{\frac{1}{\pi}}-\,\mu_\textrm{g}
\,e^{2b_\textrm{g} \phi_{B}^{cl}}
\;2b_\textrm{g}^2\;g_0\,(\,g_0+2\phi_M\,)\,
\right]\,d ^2z 
  \nonumber \\
& & 
\rule{0pt}{1cm}\hspace{-4cm}= \hspace{.2cm}
\int_{\Delta}  \left[\;\frac{1}{\pi}
  \partial_z  \phi_M \partial_{\bar{z}}\phi_M +
  \,\frac{2}{\pi}\,\frac{\phi_M^2}{(\,1-z\bar{z}\,)^2}\;
  \right]\,d ^2z\,
 +   \sum_{k=3}^{\infty}\,\frac{1}{k!}\, 
\int_{\Delta} \,\frac{\big(\,2b_\textrm{g}
(\,\phi_M+g_0\,)\, \big)^k}{\pi b_\textrm{g}^2(\,1-z\bar{z}\,)^2}\;
d ^2z \nonumber \\
& &  
\rule{0pt}{.8cm}\hspace{-4cm}= \hspace{.2cm}
S_{\Delta,0,M}[ \,\phi_M,\phi_B^{cl}\,]+
\widetilde{S}_{\Delta,N,M}[ \,\phi_M,\phi_B^{cl}\,] 
\end{eqnarray}
where
\begin{equation}
 \label{standard action}
S_{\Delta,0,M}[ \,\phi_M,\phi_B^{cl}\,]=
\int_{\Delta}  
\left[\;\frac{1}{\pi} \,\partial_z \phi_M \,\partial_{\bar{z}}\phi_M  
+ \frac{e^{2b_\textrm{g}\phi_M}-1-2b_\textrm{g}\phi_M}{\pi b_\textrm{g}^2\,
(\,1-z\bar{z}\,)^2}\;\right]\,d ^2z 
\end{equation}
is formally identical to the action for the quantum field of
the standard approach.\\
The source field $g_0(z; z_1, \dots z_N)$ is contained only in 
$\widetilde{S}_{\Delta,N,M}[ \,\phi_M,\phi_B^{cl}\,]$
\begin{eqnarray}
 \label{action for source field}
\hspace{-1cm}
\widetilde{S}_{\Delta,N,M}[ \,\phi_M,\phi_B^{cl}\,] & = &
\frac{2}{b_\textrm{g}}\,\int_{\Delta}\frac{g_0}{\pi \,(\,1-z\bar{z}\,)^2}\,
\left[\,e^{2b_\textrm{g}\phi_M}-1-2b_\textrm{g}\phi_M\,\right]\,d^2 z
\nonumber \\
& & +\,2\,\int_{\Delta}\frac{g_0^2}{\pi \,(\,1-z\bar{z}\,)^2}\,
\left[\,e^{2b_\textrm{g}\phi_M}-1\,\right]\,d^2 z\nonumber \\
& & +\,
\frac{1}{b_\textrm{g}^2}\,\sum_{k=3}^{\infty}\frac{(\,2b_\textrm{g}\,)^k}{k!}\,
\,\int_{\Delta}\frac{g_0^k}{\pi \,(\,1-z\bar{z}\,)^2}\;
\left[\,e^{2b_\textrm{g}\phi_M}\,\right]\,d^2 z~. 
\end{eqnarray}
 From the second form of (\ref{action phiM N=1}), we see that 
the propagator $g(z,z')=\langle \,\phi_M(z)\phi_M(z')\,\rangle$
in the geometric approach is the same as in the standard approach 
\cite{ZZ:Pseudosphere}.\\
In order to compare the geometric approach with the standard one,
the third form of (\ref{action phiM N=1}) turns out to be the
most useful.\\

\section{One point function}
 \label{One point function}

In this section we shall provide the perturbative expansion of the one point
function on the pseudosphere in the geometric approach and we shall compare it
with the results of the standard approach 
\cite{ZZ:Pseudosphere, tetrahedron}.\\
We recall that, within the geometric approach, the
expectation value of the Liouville vertex operator
$V_{\alpha_1}(z_1)=e^{2\alpha_1\phi(z_1)}$ is given by
\begin{equation}
 \label{geometric one-point function}
\left\langle \, V_{\alpha_1}(z_1) \,\right\rangle=
\left\langle \, e^{2\alpha_1 \phi(z_1)}\,\right\rangle 
 \,=\,\frac{\displaystyle \int_{\mathcal{C}(\Delta)}
\mathcal{D}\,[\, \phi \,]\,\, 
e^{-S_{\Delta,\,1} 
\left[ \,\phi \,\right] }}{\displaystyle \int_{\mathcal{C}(\Delta)}
\mathcal{D}\,[\, \phi \,]\,\, 
e^{-S_{\Delta,\,0} 
\left[ \,\phi \,\right] }} 
\end{equation} 
being $S_{\Delta,1}\left[ \,\phi \,\right]$ the geometric action
(\ref{geometric action pseudosphere phiM}) with one source 
of charge $\alpha_1$.\\
To perform a perturbative calculation, we choose
$\phi_B=\phi_B^{cl}$ given in (\ref{phiBcl pseudosphere}),
as done in \cite{ZZ:Pseudosphere}.\\
Then the geometric action (\ref{geometric action pseudosphere phiM}) with
one source simplifies to
\begin{eqnarray} 
 \label{geometric action pseudosphere phiM N=1 phiBcl}
S_{\Delta,1}[ \,\phi\,]  
& = & 
S_{\Delta,B}[ \,\phi_B^{cl}\,]+ S_{\Delta,1,M}[ \,\phi_M,\phi_B^{cl}\,] 
\nonumber \\
& & \rule{0pt}{.7cm}+\,\frac{\alpha_1}{b_\textrm{g}}\,\log
\left[\,\pi b_\textrm{g}^2 \mu \,(\,1-z_1\bar{z}_1\,)^2 \, \right]\,-\,
\alpha_1^2\,\left[\,\log\,(\,1-z_1\bar{z}_1\,)^2-2\,\right]
\end{eqnarray}
where $S_{\Delta,1,M}[ \,\phi_M,\phi_B^{cl}\,] $ is the action 
(\ref{action phiM N=1}) with $N=1$
and the source field is given by $g_0(z;z_1)=2\alpha_1 g(z,z_1)$.\\ 
The one point function (\ref{geometric one-point function})
in the geometric approach can be written as
\begin{equation}
 \label{geometric one-point phiM}
\left\langle \,V_{\alpha_1}(z_1) \,\right\rangle =
\frac{U_\textrm{g}}{(\,1-z_1\bar{z}_1\,)^{\,2\alpha_1(\,Q-\alpha_1\,)}}
\end{equation}
where $Q=1/b_\textrm{g}$ and
\begin{eqnarray}
 \label{U}
U_\textrm{g} &=&\,
\left[\,\pi b_\textrm{g}^2 \mu_\textrm{g} \,\right]^{-\alpha_1 /b_\textrm{g}}
e^{-2\alpha_1^2}\, 
\frac{\displaystyle \int_{\mathcal{C}  (\Delta)}
  \mathcal{D}\,[\,\phi_M\,]\, 
e^{- S_{\Delta,1,M}[ \,\phi_M,\phi_B^{cl}\,]}}{\displaystyle
  \int_{\mathcal{C}  (\Delta)}
  \mathcal{D}\,[\,\phi_M\,]\, 
e^{- S_{\Delta,0,M}[ \,\phi_M,\phi_B^{cl}\,]}} \nonumber \\
&=& \rule{0pt}{.8cm}
\,\left[\,\pi b_\textrm{g}^2 \mu_\textrm{g} \,\right]^{-\alpha_1/b_\textrm{g}}
e^{-2\alpha_1^2}\,
\left\langle\,e^{-\widetilde{S}_{\Delta,1,M}}\,\right\rangle_{z_1}~.
\end{eqnarray}
We observe here that $U_\textrm{g}$, 
taken as a function of the charge $\alpha_1$, has the
correct normalization $U_\textrm{g}(\alpha_1=0)=1$ \cite{ZZ:Pseudosphere},
being $\widetilde{S}_{\Delta,1,M}=0$ when $\alpha_1=0$.
Moreover, the mean value in the second line of (\ref{U}) does not 
depend on $\mu_\textrm{g}$.\\
As noticed in  \cite{Takhtajan:Equivalence}, one obtains the
same central charge and the same quantum conformal dimensions 
of the standard approach by relating
the coupling constant of the geometric approach $b_\textrm{g}$ to the coupling
constant of the standard approach $b$ as follows
\begin{equation}
 \label{Qgeometric=Qstandard}
\frac{1}{b_\textrm{g}}=\frac{1}{b}+b~.
\end{equation}
Thus, the comparison between the one point functions of the two
theories reduces to the comparison between 
$U_\textrm{g}(\alpha_1,b_\textrm{g},\mu_\textrm{g})$ in the
geometric approach and $U(\alpha_1,b,\mu)$ in the standard approach
\cite{ZZ:Pseudosphere}.\\
Following \cite{ZZ:Pseudosphere}, it is useful to consider, instead
of $\left\langle \, e^{2\alpha_1\phi(z_1)}\,\right\rangle$, the
cumulant expansion 
\begin{eqnarray}
 \label{ln(one-point function)}
\log \left\langle \, e^{2\alpha_1\phi(z_1)}\,\right\rangle 
& = &
\sum_{n=1}^\infty \frac{(\,2\alpha_1)^n}{n!}\;G_n^{\textrm{g}}   \nonumber \\
& = & 2\alpha_1\,
\left[\,-\,\frac{1}{2b_\textrm{g}}\,\log\,(\,1-z_1\bar{z}_1\,)^2 \,\right]
+\,\frac{(\,2\alpha_1\,)^2}{2}\,\Big[\,\log\,(\,1-z_1\bar{z}_1\,) \;\Big]
+\log\,U_\textrm{g}~.
\nonumber \\
& &
\end{eqnarray}
Using the expression of $U_\textrm{g}$ given in (\ref{U}), we get
\begin{eqnarray}
 \label{G1}
G_1^{\textrm{g}}  &=& \langle \,\phi(z_1) \,\rangle 
=
-\,\frac{1}{2b_\textrm{g}}
\log\,(\,1-z_1\bar{z}_1\,)^2 -\frac{1}{2b_\textrm{g}}\log\,[\,\pi
 b_\textrm{g}^2 \mu_\textrm{g}\,]\, -\,\frac{1}{2}\,
\left\langle\, \widetilde{S}_{\Delta,1,M}^{(1)}\,\right\rangle  \\
& & \nonumber \\
 \label{G2}
G_2^{\textrm{g}} & = & 
\langle\, \phi^2(z_1) \,\rangle -\langle\,\phi(z_1)\,\rangle^2 \nonumber \\   
&=&  \log\,(\,1-z_1\bar{z}_1\,)\,-1+\,\frac{1}{2^2}\,
\Bigg\{\,-\,\left\langle\, \widetilde{S}_{\Delta,1,M}^{(2)}\,\right\rangle 
\,+\,\Big\langle \left( \widetilde{S}_{\Delta,1,M}^{(1)}\right)^2\Big\rangle\,
\,-\,
\left(\,\Big\langle\,\widetilde{S}_{\Delta,1,M}^{(1)}\Big\rangle\,\right)^2\,
\Bigg\}\nonumber \\
& &  \\
 \label{G3}
G_3^{\textrm{g}} & = & \langle\, \phi^3(z_1) \,\rangle -
              3 \,\langle\, \phi^2(z_1)
              \,\rangle\,\langle\,\phi(z_1)\,\rangle+ 
              2 \,\langle\, \phi(z_1) \,\rangle^3  \nonumber \\
&=&\,\frac{1}{2^3}\,\rule{0pt}{1.1cm}
\Bigg\{\,-\,\left\langle\, \widetilde{S}_{\Delta,1,M}^{(3)}\,\right\rangle 
\,+3\, \Big\langle\,\widetilde{S}_{\Delta,1,M}^{(1)}\,
\widetilde{S}_{\Delta,1,M}^{(2)}\,\Big\rangle\,-
3\, \Big\langle\,\widetilde{S}_{\Delta,1,M}^{(1)}\,\Big\rangle\,
\Big\langle\,\widetilde{S}_{\Delta,1,M}^{(2)}\,\Big\rangle\,\nonumber \\
& & \hspace{1.05cm}\,-\;
\Big\langle\left(\,\widetilde{S}_{\Delta,1,M}^{(1)}\,\right)^3\Big\rangle\,
+3\, \Big\langle\,\widetilde{S}_{\Delta,1,M}^{(1)}\,\Big\rangle\,
\Big\langle\left(\,\widetilde{S}_{\Delta,1,M}^{(1)}\,\right)^2\Big\rangle\,-\, 
2\,
\left(\,\Big\langle\,\widetilde{S}_{\Delta,1,M}^{(1)}\Big\rangle\,\right)^3\,
\,\Bigg\}\nonumber \\
& &  
\end{eqnarray}
where 
\begin{equation}
\widetilde{S}_{\Delta,1,M}^{(k)}=\left.\frac{\partial^k}{\partial \alpha_1^k}\,
\widetilde{S}_{\Delta,1,M}\,\right|_{\alpha_1=0}
\end{equation}
and the mean values are taken with respect to the action 
(\ref{standard action}).\\
Using (\ref{G1}) and considering the $O(\alpha_1)$ contribution
to (\ref{action for source field}), one obtains
\begin {equation}
 \label{G_1}
 G_1^{\textrm{g}} 
= \phi_B^{cl}(z_1)+\left[\;-\,\frac{1}{b_\textrm{g}} \int_{\Delta}
   \frac{g_0^{(1)}(z;z_1)}{\pi(\,1-z\bar{z}\,)^2} \left( \;\langle\,
     e^{2b_\textrm{g}\phi_M(z)}\,\rangle-1-
     2b_\textrm{g} \langle\,\phi_M(z)\,\rangle \; \right) d^2 z \;\right]
\end{equation}
where $g_0^{(1)}(z;z_1)=2\,g(z,z_1)$ is 
the derivative with respect to $\alpha_1$ of the classical 
source field $g_0(z;z_1)$.\\
The graphs contributing to $G_1^{\textrm{g}}$ up to $O(b_\textrm{g}^4)$ 
included are shown below, where the dashed lines represent 
$g_0^{(1)}(z;z_1)$.\\
\begin{equation}
 \label{G1geomgraphs}
\begin{array}{lll}
\vspace{1cm}
G_1^{\textrm{g}}&=& \phi_B^{cl}(z_1)\hspace{.2cm} +\hspace{.2cm}
b_\textrm{g}\;\left[\; \hspace{.4cm} 
\begin{minipage}[c]{3.2cm} 
  \includegraphics[width=2.4cm]{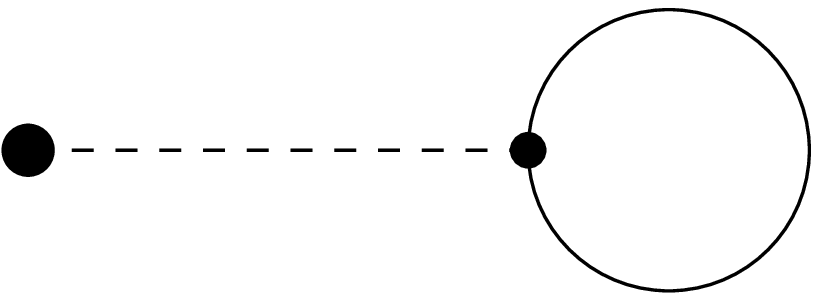}
\end{minipage} 
 \hspace{-.1cm} \right] 
  \\ 
\vspace{.4cm}
  & &
+ \hspace{.2cm} b_\textrm{g}^3\;\left[ \hspace{.4cm}
\begin{array}{ccc}
\vspace{.5cm}
\hspace{-1.05cm}
\begin{minipage}[c]{2.5cm} 
  \includegraphics[width=2.7cm]{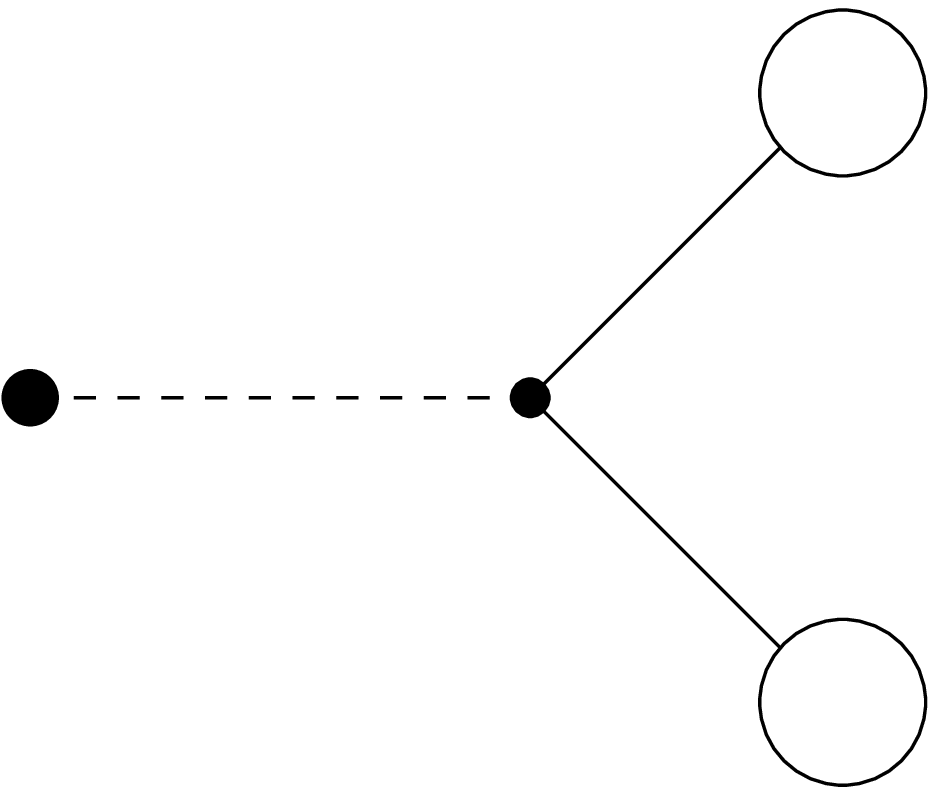}
\end{minipage} 
&
\begin{minipage}[c]{2.5cm} 
  \includegraphics[width=2.9cm]{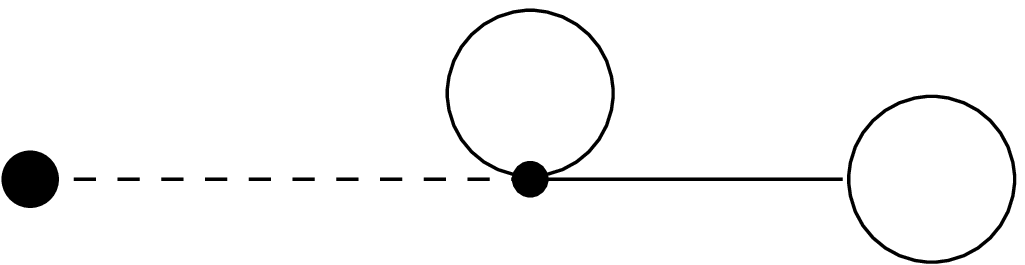}
\end{minipage} 
&
\hspace{.8cm}
\begin{minipage}[c]{2.5cm} 
  \includegraphics[width=1.7cm]{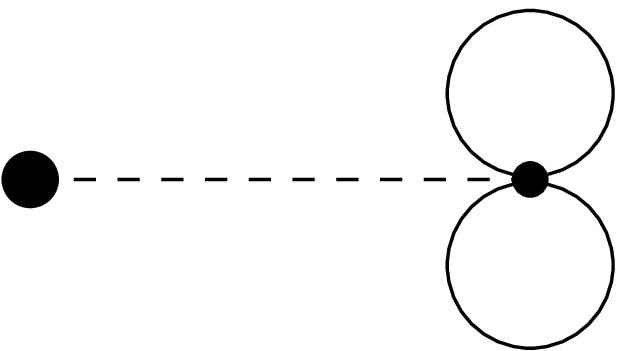}
\end{minipage} 
\\
\vspace{.7cm}
\begin{minipage}[c]{3.5cm} 
  \includegraphics[width=3.6cm]{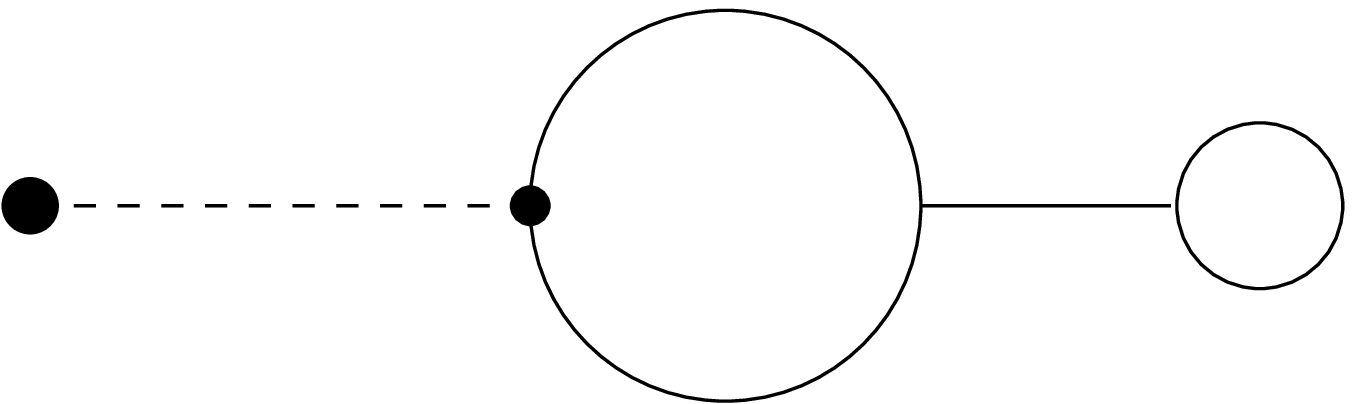}
\end{minipage} 
&
\hspace{.95cm}
\begin{minipage}[c]{3.5cm} 
  \includegraphics[width=3.1cm]{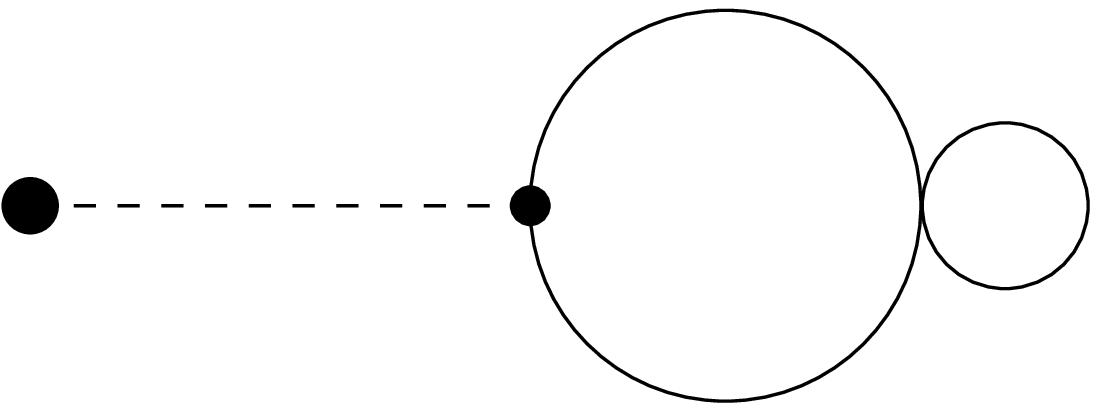}
\end{minipage} 
&  \\
\begin{minipage}[c]{3.5cm} 
  \includegraphics[width=2.9cm]{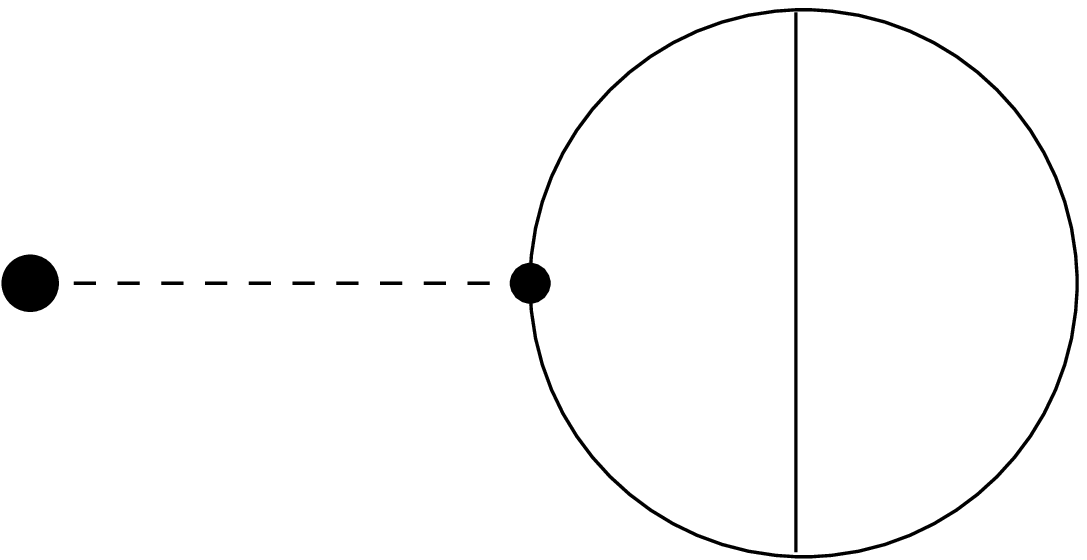}
\end{minipage} 
&
\hspace{.9cm}
\begin{minipage}[c]{3.5cm} 
  \includegraphics[width=2.9cm]{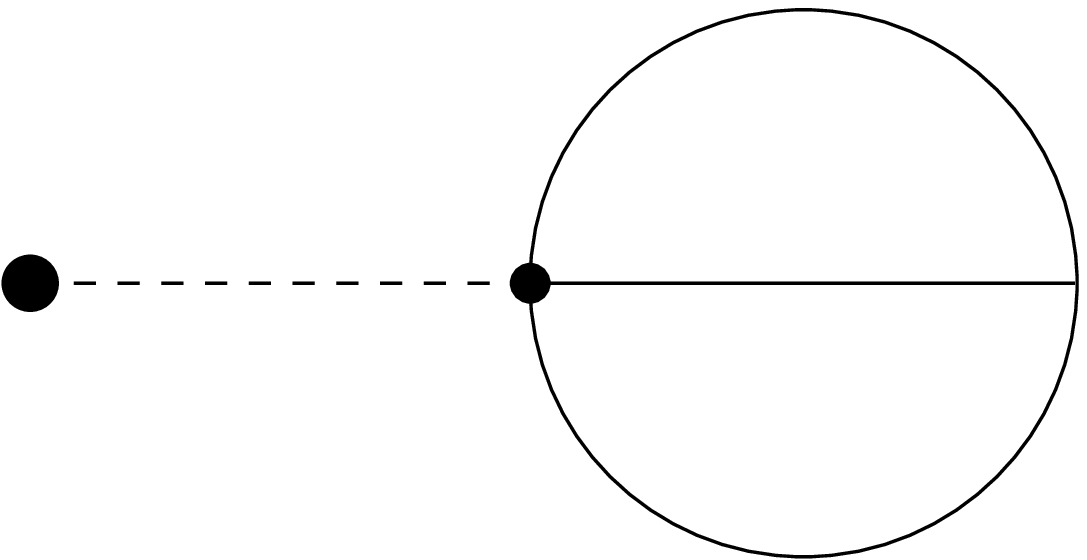}
\end{minipage} 
& 
\end{array}
 \hspace{0cm}\right] \\
& & + \hspace{.2cm} O(b_\textrm{g}^5)~.
\end{array}
\end{equation}

\phantom{sggkgkjhgajvaa}

\noindent In the perturbative expansion divergent graphs appear 
due to the occurrence of $g(z,z)$; therefore the
theory has to be regulated.\\
The proposal of  \cite{Takhtajan:Topics} is to regulate such a
propagator at coincident points with the Hadamard procedure
(\,see also \cite{Hadamard, Garabedian, Moretti}\,), which
amounts to set
\begin{equation}
g(z,z) \equiv \lim_{z' \rightarrow z}
\left[\;g(z,z')+\frac{1}{2}\log\eta +\log 2 \;\right]~. 
\end{equation}
This limit gives $g(z,z)=-1+\log 2$; but for sake of
generality we shall consider $g(z,z)=C$.\\
We notice that within the geometric approach $g(z,z)$ has to be 
a constant in $z$, which implies  $U_\textrm{g}$  constant in $z_1$,
otherwise the relation between the central charge
\begin{equation}
 \label{geometric central charge}
c_\textrm{g}=1+6 \left(\,\frac{1}{b_\textrm{g}}\,\right)^2
\end{equation}
and the quantum conformal dimensions of 
$\left\langle \,e^{2\alpha_1\phi(z_1)}\,\right\rangle$, which are
\begin{equation}
 \label{quantumgeomdimensions}
\Delta_{\alpha_1} = 
\alpha_1\,\left(\,\frac{1}{b_\textrm{g}}-\alpha_1\,\right)
\end{equation}
would be violated.\\
As already noticed in \cite{Takhtajan:Equivalence}, expression
(\ref{quantumgeomdimensions}) provides the cosmological term 
with quantum conformal dimensions 
$(\,1-b_\textrm{g}^2,1-b_\textrm{g}^2\,)$.\\
The $O(b_\textrm{g})$ contribution to $G_1^\textrm{g}$ 
is given by $-\,b_\textrm{g}\,C$.
Concerning the $O(b_\textrm{g}^3)$ contribution, 
we find that the three graphs contained 
in the second line of (\ref{G1geomgraphs}) sum up to zero and the same 
happens for the two graphs of the third line. The whole 
$O(b_\textrm{g}^3)$ contribution comes from the two graphs of the fourth line:
their values are $(\pi^2-15)/18$ and $2/3$ respectively.\\ 
Thus, the expansion of $G_1^\textrm{g}$ in the coupling constant 
$b_\textrm{g}$ within the geometric approach,
up to $O(b_\textrm{g}^4)$ included, is 
\begin{equation}
 \label{G_1 value}
G_1^\textrm{g}= -\,\frac{1}{2b_\textrm{g}}\,\log
\left[\,\pi b_\textrm{g}^2 \mu_\textrm{g}\,(\,1-z_1\bar{z}_1\,)^2\,\right]
-b_\textrm{g}\, C + 
b_\textrm{g}^3 \,\Bigg(\,\frac{\pi^2}{18}\,-\,\frac{1}{6}\;\Bigg)
\,+ O(b_\textrm{g}^5)~.
\end{equation}
Notice that, because of the cancellation between graphs 
explained just above, the regulator $C$ appears only in the contribution 
$O(b_\textrm{g})$.
The contribution of the term in the square
brackets of (\ref{G_1}) is given by $\langle\,\phi_M(z_1)\,\rangle$.
Indeed, varying $\phi_M$ in (\ref{standard action}), 
we obtain for the expectation value of the equation of motion
\begin{equation}
 \label{<eq. motion phiM> with a=0}
-\,\frac{1}{\pi}\, \partial_z \partial_{\bar{z}} \,\langle\,\phi_M\,\rangle +
\mu_\textrm{g} b_\textrm{g} e^{2b_\textrm{g}\phi_B^{cl}} 
\left(\,\langle\,e^{2b_\textrm{g}\phi_M}\,\rangle -1\,\right)
\,=0
\end{equation}
which can be rewritten, using the equation of the propagator 
and the expression (\ref{phiBcl pseudosphere}), as follows
\begin{equation}
 \label{Ward Identity}
\langle \,\phi_M(z_1) \,\rangle \;=\;-\,\frac{2}{b_\textrm{g}}  \int_{\Delta} 
       \frac{g(z_1,z)}{\pi(\,1-z\bar{z}\,)^2}
       \left( \;\langle \,e^{2b_\textrm{g}\phi_M(z)}\,\rangle-1-
       2b_\textrm{g} \langle\,\phi_M(z)\,\rangle \; \right) d^2 z~.
\end{equation}
Being $g_0^{(1)}(z;z_1)=2\,g(z,z_1)$, we
recognize in the term in square brackets of (\ref{G_1})
the r.h.s. of the previous Ward identity.\\
Equation (\ref{Ward Identity}) has the same form of the 
Ward identity obtained by ZZ
within the standard approach \cite{ZZ:Pseudosphere} 
for the quantum field $\chi$ and verified up to $O(b^3)$ 
with their regulator.\\
Direct computation of $\langle \,\phi_M(z_1) \,\rangle $ with the $C$
(\,Hadamard\,) 
regulator up to $b_\textrm{g}^4$ included agrees with the r.h.s. of 
(\ref{G_1 value}). 
On the other hand, we have checked the validity of the equation of motion 
(\ref{<eq. motion phiM> with a=0}) in the $C$
regularized theory up to $b_\textrm{g}^4$ included. 
Indeed, (\ref{Ward Identity}) follows directly from the independence 
of $\langle\,\phi_M(z)\,\rangle$ on $z$ and 
from $\langle\,e^{2b_\textrm{g}\phi_M}\,\rangle =1$.
The perturbative expansion (\ref{G_1 value}) gives the constancy of
$\langle \,\phi_M(z) \,\rangle $ up to $b_\textrm{g}^4$ included, while
we have checked that, with the $C$ regulator, 
$\langle\,e^{2b_\textrm{g}\phi_M}\,\rangle =1$ is valid up to 
$b^5_\textrm{g}$ included.
We expect such an identity to be valid to all orders in the coupling 
constant $b_\textrm{g}$.\\
Therefore we obtain up to $b^4_\textrm{g}$ included
\begin{equation}
 \label{G_1=phiB+<phiM>}
G_1^\textrm{g}
=\langle \,\phi(z_1)\,\rangle
= \phi_B^{cl}(z_1)+\langle\,\phi_M(z_1)\,\rangle~.  
\end{equation}
Notice that, due to (\ref{Ward Identity}), within the geometric
approach  $\langle\,\phi_M(z_1)\,\rangle$ cannot depend 
on the position $z_1$ if we want to keep for $Q$ the value 
$1/b_\textrm{g}$.
As seen above this requirement is satisfied by using the $SU(1,1)$
invariant regulator $C$
\cite{Takhtajan:Topics}.\\
The structure (\ref{G_1=phiB+<phiM>}) was obtained also in the
standard approach by \cite{ZZ:Pseudosphere}, with the difference that
$\langle\,\phi_M(z_1)\,\rangle $ is not $z_1$
independent because of the different choice of the regulator.  In the
standard approach that choice of regulator is necessary in order to
provide the correct quantum conformal dimensions of the hamiltonian treatment
\cite{CT}
\begin{equation}
 \label{standard anomalous dimensions}
\Delta_{\alpha_1}= \alpha_1\, \left( \,\frac{1}{b}+b-\alpha_1\, \right)~.
\end{equation}
Dependence on $z_1$ of $G_1^\textrm{g}$ given in (\ref{G_1 value}) 
compared to that of the standard approach \cite{ZZ:Pseudosphere} imposes 
the relation (\ref{Qgeometric=Qstandard}) between $b_\textrm{g}$ and $b$,
while the agreement between the constant terms 
can be obtained by choosing $\mu_\textrm{g}$ as a
proper function of $\mu$, $b$ and $C$.\\
Concerning $G_2^\textrm{g}$, from (\ref{G2}) we get
\begin {eqnarray}
 \label{G_2} 
G_2^\textrm{g} 
 &=&\rule{0pt}{.3cm}
 \log\,(\,1-z_1\bar{z}_1\,)-1
 -\,\int_{\Delta}
 \frac{(\,g_0^{(1)}(z;z_1)\,)^2}{\pi\,(\,1-z\bar{z}\,)^2} \left( \;\langle\,
   e^{2b_\textrm{g}\phi_M(z)}\,\rangle-1\; \right) d^2 z \nonumber \\
 & & \rule{0pt}{1cm}
 +\, \frac{1}{b_\textrm{g}^2} \int_{\Delta}
 \frac{g_0^{(1)}(z;z_1)}{\pi(\,1-z\bar{z}\,)^2}\;
 \frac{g_0^{(1)}(z';z_1)}{\pi(\,1-z'\bar{z}'\,)^2} \;\langle \,\Big(
   \;e^{2b_\textrm{g}\phi_M(z)}-1-2b_\textrm{g} \phi_M(z)\; \Big)\; 
 \nonumber \\
 & & 
\hspace{6.5cm} 
\Big(\;e^{2b_\textrm{g}\phi_M(z')}-1-2b_\textrm{g}\phi_M(z')\;\Big)\,\rangle\,
 d^2 z \,d^2 z' \nonumber \\
 & &
-\,\left( \,-\,\frac{1}{b_\textrm{g}} \int_{\Delta}
\frac{g_0^{(1)}(z;z_1)}{\pi(\,1-z\bar{z}\,)^2} 
\left( \;\langle\,e^{2b_\textrm{g}\phi_M(z)}\,\rangle-1-
2b_\textrm{g}\langle\,\phi_M(z)\,\rangle \; \right) d^2 z \,\right)^2~.
\nonumber\\
 & &
\end{eqnarray}
Through the Ward identity (\ref{Ward Identity}),
the last term reduces to $\langle\,\phi_M(z_1)\,\rangle^2$,
which has already been computed.
The graphs that contribute to $G_2^\textrm{g}$  are given below\\
\begin{equation}
\begin{array}{lll}
\vspace{.5cm}
G_2^\textrm{g} & = & \log\,(\,1-z_1\bar{z}_1\,)\,-1\\ 
\vspace{.4cm}
& & + \hspace{.2cm} 
b^2_\textrm{g}\;\left[ \hspace{.4cm}\rule{0pt}{2.8cm}
\begin{array}{ccc}
\vspace{.7cm}
\hspace{-.0cm}
\begin{minipage}[c]{3.5cm} 
  \includegraphics[width=3cm]{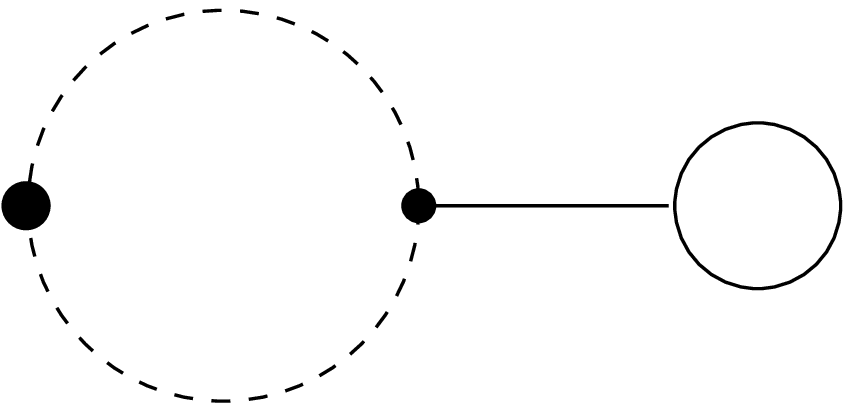}
\end{minipage} 
&
\hspace{.3cm}
\begin{minipage}[c]{3.5cm} 
  \includegraphics[width=2cm]{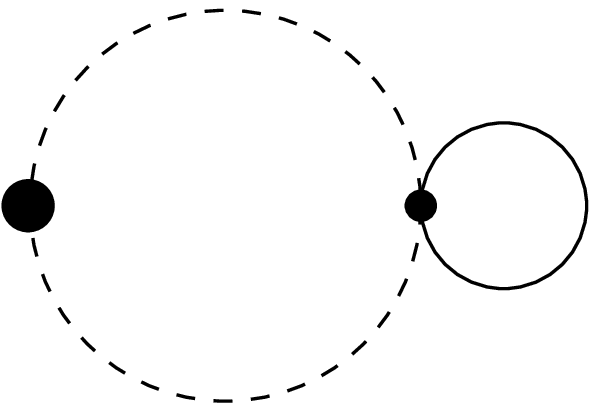}
\end{minipage} 
& \\
\begin{minipage}[c]{3.5cm} 
  \includegraphics[width=2cm]{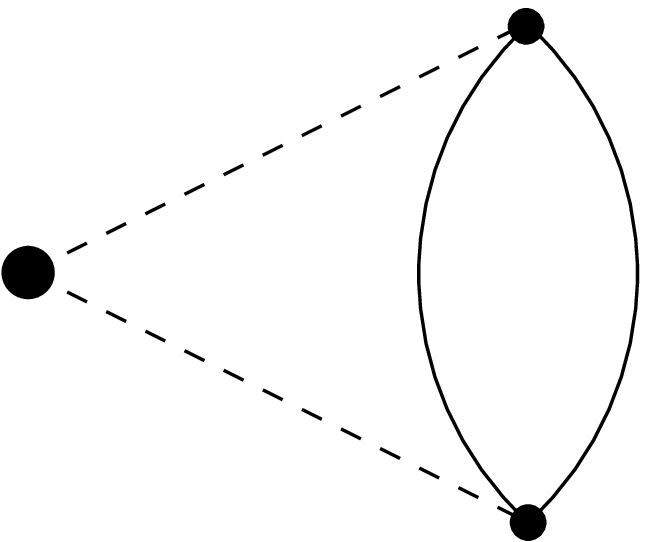}
\end{minipage}
& \hspace{.5cm}
\begin{minipage}[c]{3.5cm} 
  \includegraphics[width=2.3cm]{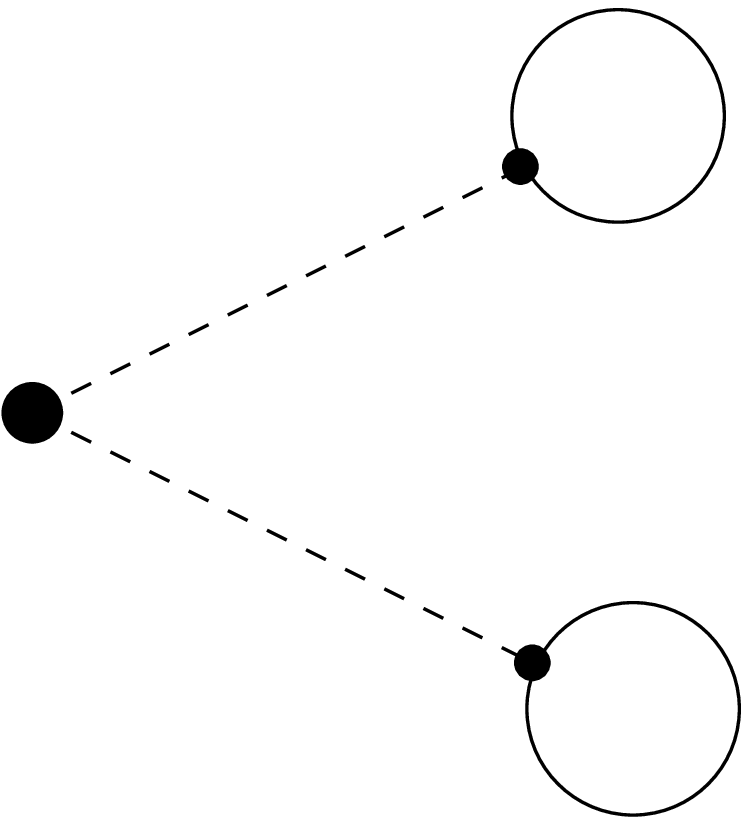}
\end{minipage} 
& \hspace{-.5cm}
\left( \hspace{.3cm}\begin{minipage}[c]{3.2cm} 
                     \includegraphics[width=2.4cm]{G1-1Ageom.eps}
                    \end{minipage} \hspace{-.4cm} \right)^2
\end{array}
 \hspace{.2cm}\right]\\
& &+ \hspace{.2cm} O(b_\textrm{g}^4)~.
\end{array}
\end{equation}
The first integral in (\ref{G_2}) gives the two graphs in the first line
of the $O(b^2_\textrm{g})$ contribution and they cancel out.
Indeed, it can be shown that this integral vanishes at least 
up to the order $b^5_\textrm{g}$ included, in agreement with the already 
discussed Ward identity (\ref{Ward Identity}).\\
The second integral in (\ref{G_2}) provides a nonvanishing  
$O(b_\textrm{g}^2)$  contribution through the first two graphs 
in the second line, but the second of these graphs simplifies 
with the third one, which is the order $O(b_\textrm{g}^2)$ of the last 
integral in (\ref{G_2}), i.e. $\langle\,\phi_M(z_1)\,\rangle^2$.\\
\noindent Thus
\begin{equation}
G_2^\textrm{g}=\;\log\,(\,1-z_1\bar{z}_1\,)-1
+b_\textrm{g}^2 \; \Bigg(\,\frac{5}{6}\,-\,\frac{\pi^2}{18} \,\Bigg)
+O(b_\textrm{g}^4) 
\end{equation}
and it does not depend on $C$.\\
Since $b_\textrm{g}$ is fixed by (\ref{Qgeometric=Qstandard}), 
$G_2^\textrm{g}$ disagrees with the result of the standard approach
\cite{ZZ:Pseudosphere}
\begin{equation}
G_2=\log\,(\,1-z_1\bar{z}_1\,)-1
+\,b^2\,\Bigg(\,\frac{3}{2}\,-\,\frac{\pi^2}{6}\,\Bigg)+O(b^4)~.
\end{equation}
Notice that the $O(b_\textrm{g}^0)$ contribution to
$G_2^\textrm{g}$ comes from the regularization of the geometric action,
while in the standard approach the $O(b^0)$ contribution to 
$G_2$ is the result of the ZZ regularization of the one loop graph that
contributes to this order.\\
One could modify the geometric action 
(\ref{geometric action Pseudosphere}) by
replacing the subtraction term $\sum_n \alpha_n^2 \log \varepsilon^2_n$  with
$\sum_n \alpha_n^2 \log \,(\lambda_n^2 \varepsilon^2_n)$ 
with a properly chosen $b_\textrm{g}$ dependent $\lambda_n$, 
i.e. $\lambda_n(b_\textrm{g})$.
This change would remove the discrepancy between $G_2^\textrm{g}$ and $G_2$;
but still differences remain both in the
comparison between $G_3^\textrm{g}$ and $G_3$, as we shall 
show in what follows, 
and in the comparison of the two point functions, 
as we shall discuss in the next section. \\
The graphs contributing to $G_3^\textrm{g}$ in the geometric 
approach are shown below

\begin{equation}
 \label{G3geomgraphs}
\hspace{-.4cm}
\begin{array}{lll}
\vspace{1cm}
G_3^\textrm{g}  &= & 
b_\textrm{g}\;\left[\; \hspace{.4cm} \rule{0pt}{1.2cm}
\begin{minipage}[c]{3.2cm} 
  \includegraphics[width=1.8cm]{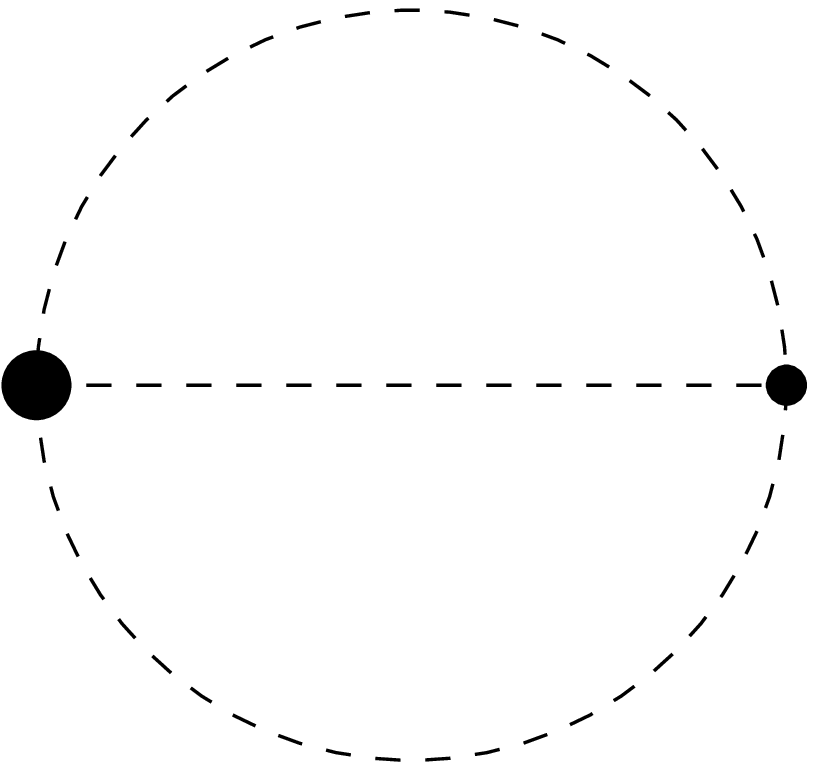}  
\end{minipage} 
 \hspace{-.8cm} \right]
  \\ 
\vspace{.8cm}
& +& b_\textrm{g}^3\;\left[ \hspace{.4cm} \rule{0pt}{4.7cm}
\begin{array}{ccc}
\vspace{1cm}
\begin{minipage}[c]{3.5cm} 
  \includegraphics[width=2.4cm]{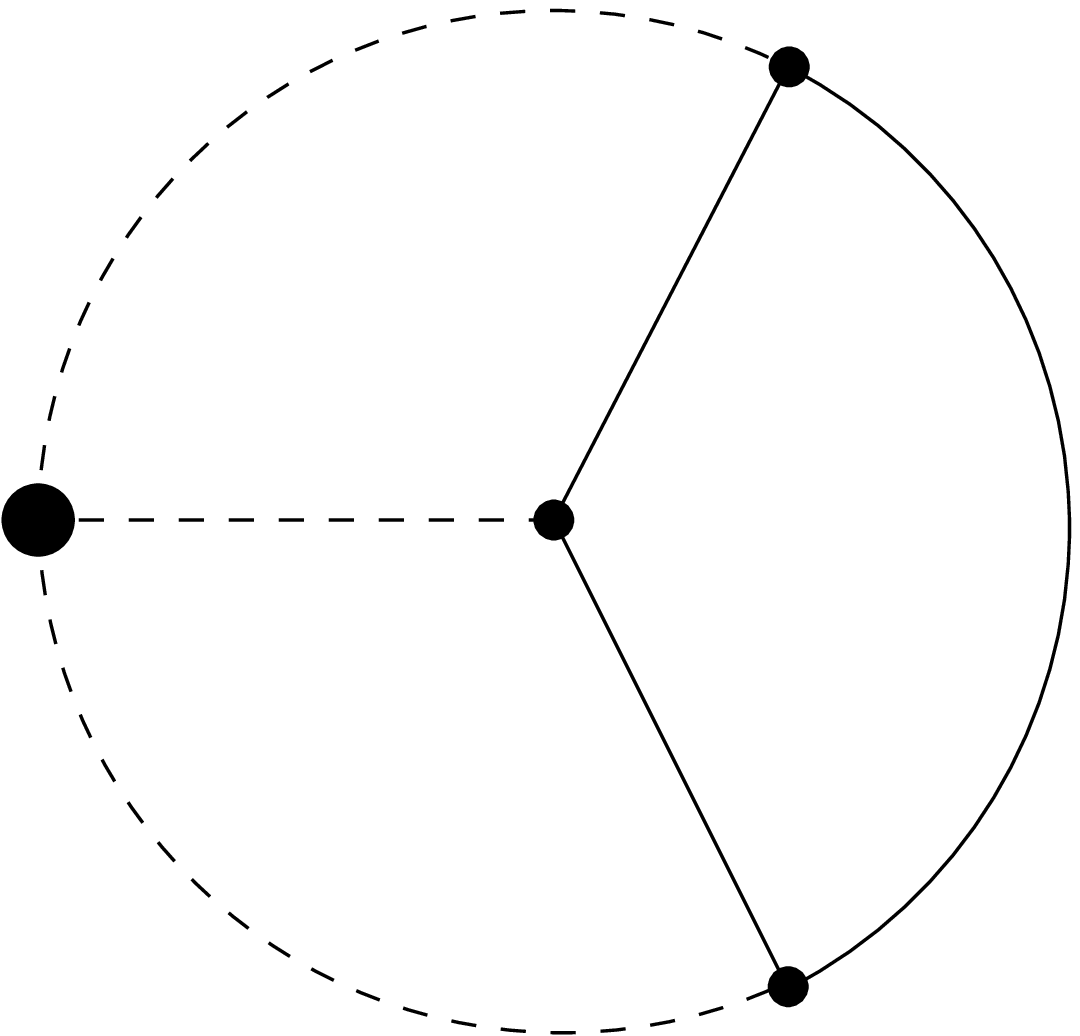}
\end{minipage} 
&
\hspace{1cm}
\begin{minipage}[c]{3.5cm} 
  \includegraphics[width=3.6cm]{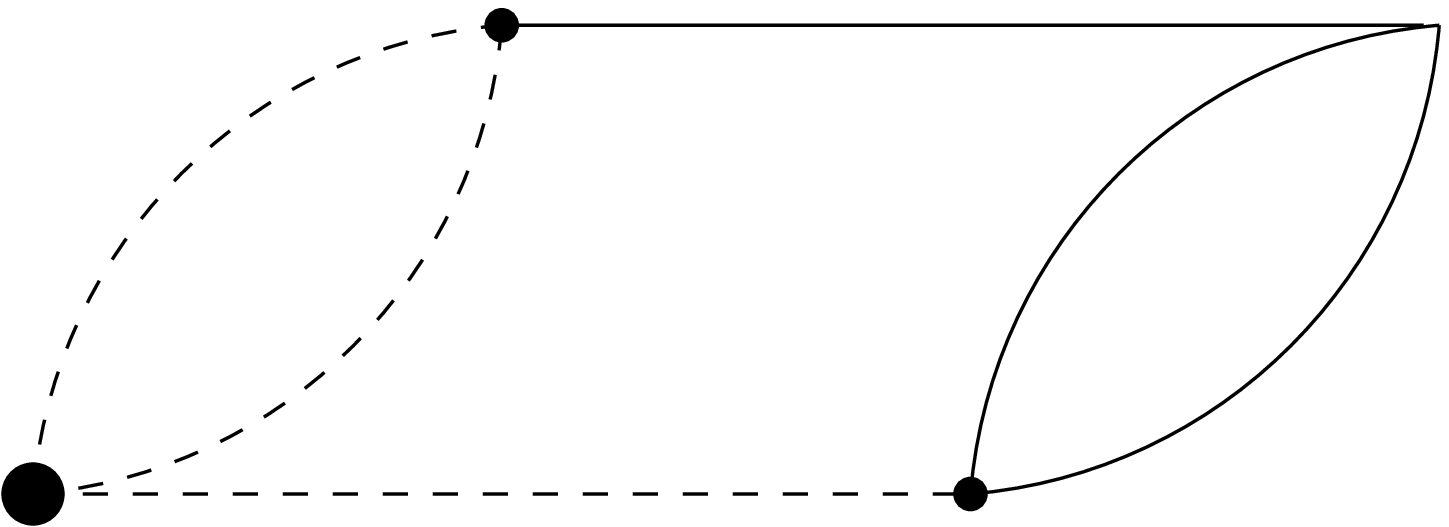}
\end{minipage} 
&
\hspace{1.4cm}
\begin{minipage}[c]{3.9cm} 
  \includegraphics[width=2.5cm]{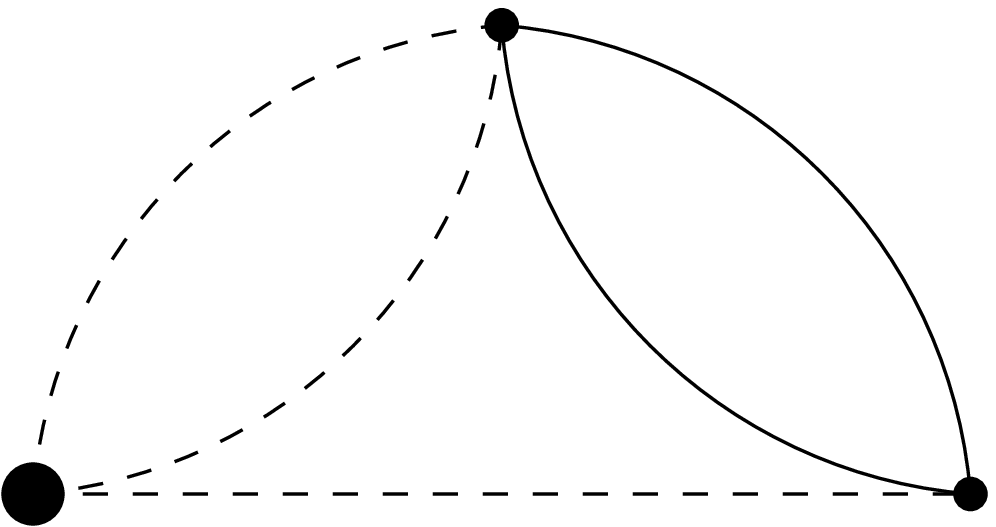}
\end{minipage} 
\\
\vspace{.7cm}
\begin{minipage}[c]{3.5cm} 
  \includegraphics[width=3.4cm]{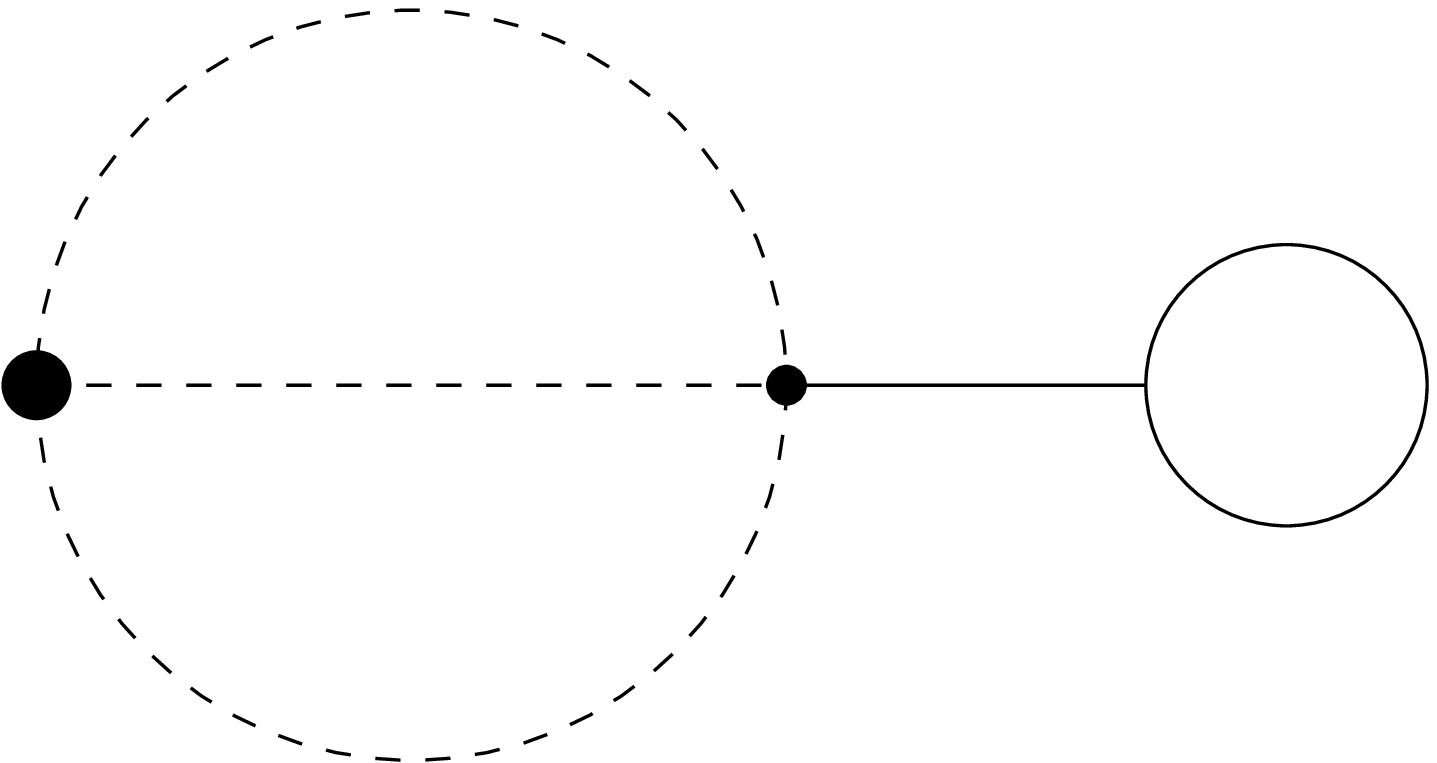}
\end{minipage} 
&
\hspace{1cm}
\begin{minipage}[c]{3.5cm} 
  \includegraphics[width=2.6cm]{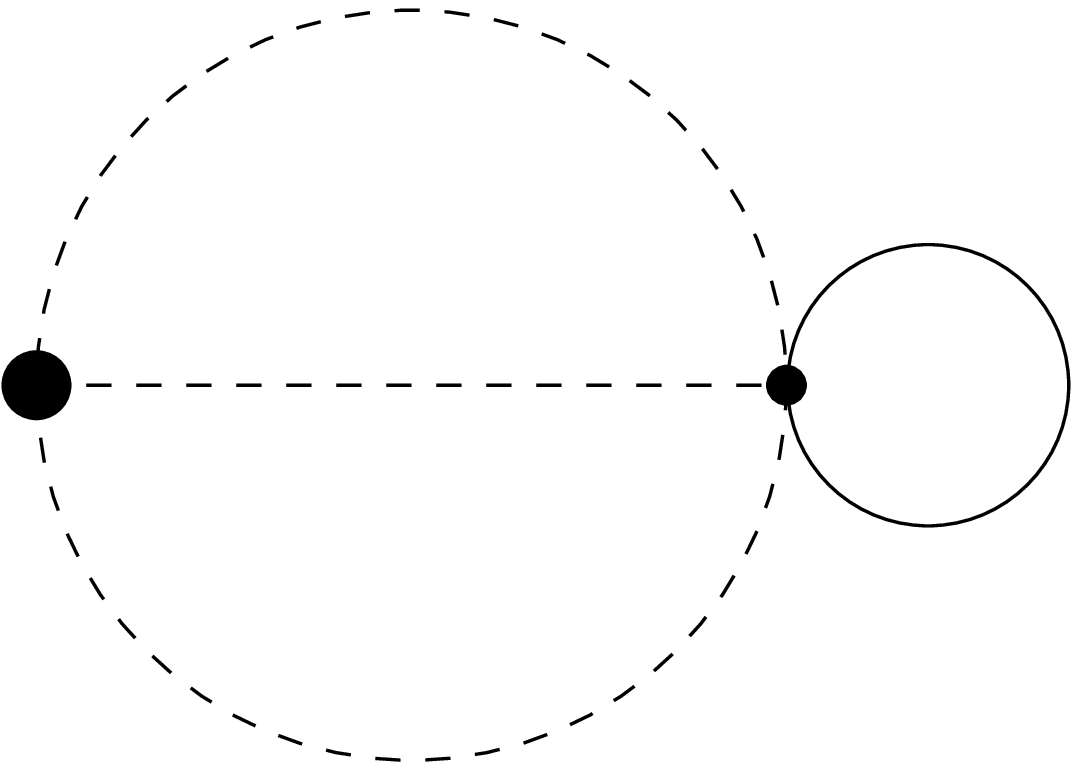}
\end{minipage} 
&
\\
\begin{minipage}[c]{3.5cm} 
  \includegraphics[width=2cm]{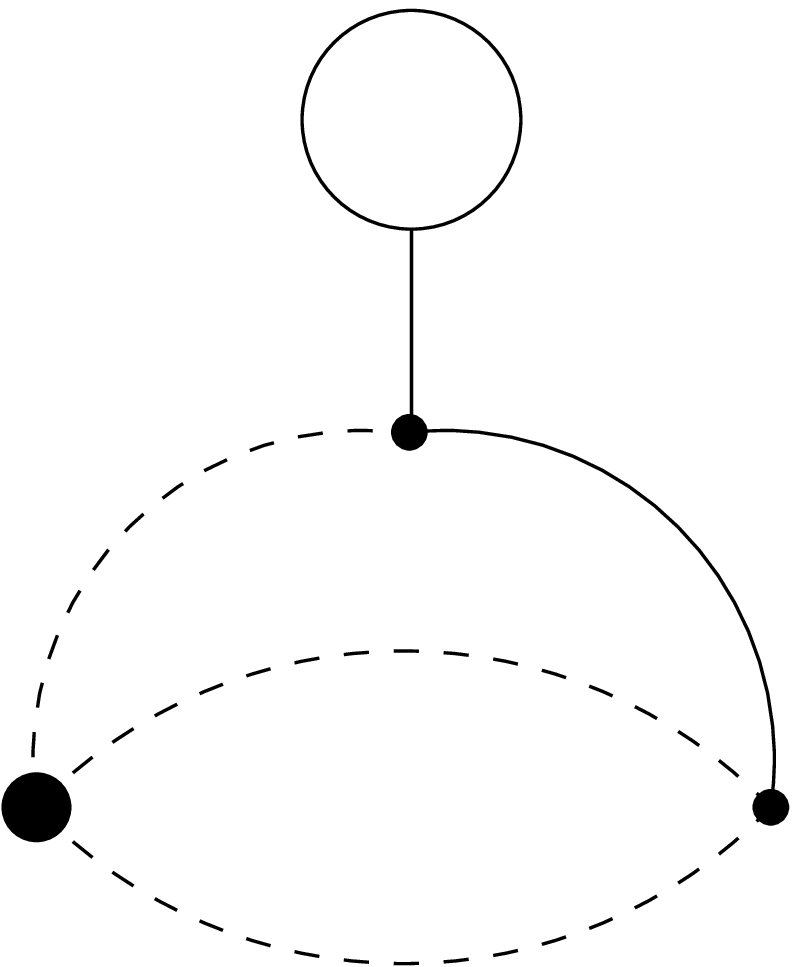}
\end{minipage} 
&
\hspace{1cm}
\begin{minipage}[c]{3.5cm} 
  \includegraphics[width=2cm]{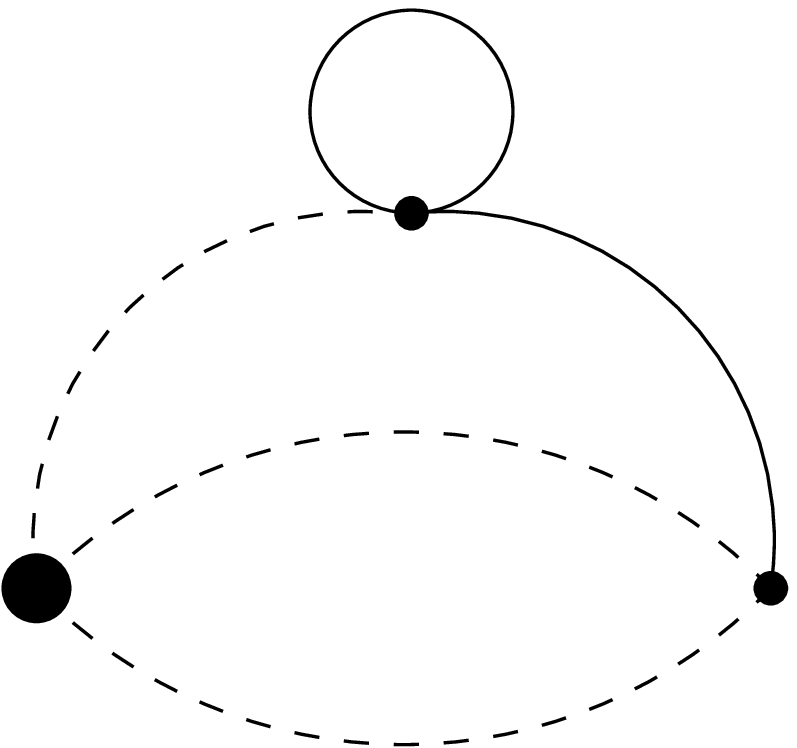}
\end{minipage} 
&
\end{array}
 \hspace{-1cm}\right] \\
&+& 
O(b_\textrm{g}^5)~.
\end{array}
\end{equation}

\rule{0pt}{.2cm}\\
\noindent 
The $O(b_\textrm{g})$ contribution to $G_3^\textrm{g}$ 
is $-\,b_\textrm{g}$.
The graphs contributing to the order $b_\textrm{g}^3$ are shown
in the big square brackets of (\ref{G3geomgraphs}). 
Adopting the Hadamard regularization, 
the two graphs in the second line sum up to zero and the same happens 
for the two ones in the third line.\\
Thus, using partially the computation given in \cite{tetrahedron},
we get the perturbative expansion of $G_3^\textrm{g}$, which is
\begin{equation}
 \label{G3geomexpansion}
G_3^\textrm{g}=-\,b_\textrm{g}\,
+\,b_\textrm{g}^3\,\Bigg(\;\frac{3}{2}\,+\,\frac{\pi^2}{6}
\,-2\,\zeta(3)\;\Bigg)\,+\,O(b_\textrm{g}^5)
\end{equation}
On the other hand, in the standard approach one gets \cite{tetrahedron}
\begin{equation}
G_3=-\,b\,+\,b^3\,\Big(\,3-2\,\zeta(3)\,\Big)\,+\,O(b^5)
\end{equation}
which is in contrast with (\ref{G3geomexpansion}), 
 taking into the account the relation (\ref{Qgeometric=Qstandard}) 
between $b$ and $b_\textrm{g}$.\\
As a further check, we have considered the first perturbative
order of the fourth cumulant.\\
Within the standard approach the $O(b^2)$ contribution to $G_4$
is given by 

\begin{equation}
G_4\hspace{.2cm}   = \hspace{.2cm}
b^2\;\left[\;\rule{0pt}{1.3cm} \hspace{.5cm} 
\begin{minipage}[c]{3.5cm} 
  \includegraphics[width=1.9cm]{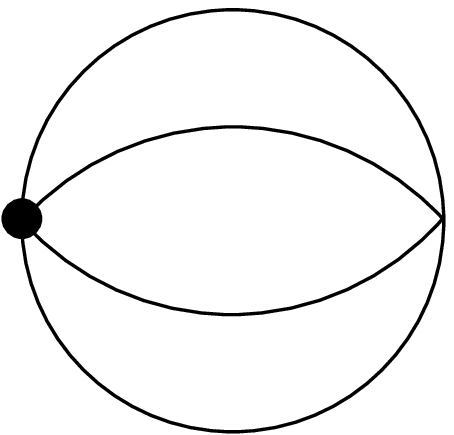}
\end{minipage} 
\hspace{-.2cm}
\begin{minipage}[c]{3.5cm} 
  \includegraphics[width=1.9cm]{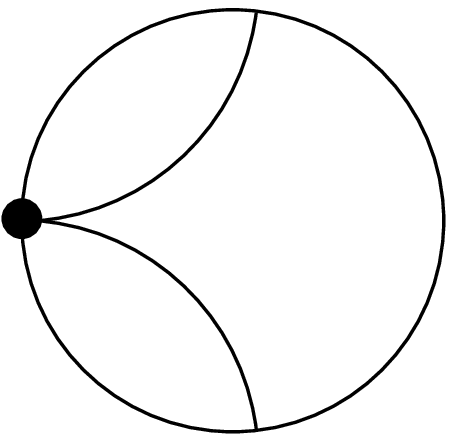}
\end{minipage} 
\hspace{-.8cm}\right] 
\hspace{.1cm} +\hspace{.1cm}  O(b^4)\;=\;-\,2\,b^2\,+ O(b^4)
\end{equation}

\mbox{ }\\
and it agrees with 
the perturbative expansion of  $G_4^\textrm{g}$ up to this order.\\
Instead of regulating the theory by giving a finite value 
to $g(z,z)$, one can proceed through the usual Pauli-Villars 
regulator technique, i.e. by introducing a regulator field $\zeta$ 
and replacing $\phi$ with $\phi+\zeta$ in the interaction lagrangian.
We show here that the $C$ and ZZ regulators arise from two
Pauli-Villars regulators whose lagrangians possess different
transformation properties.\\
Let us consider the following quadratic lagrangian
\begin{equation}
-\,\frac{1}{\pi}\,\partial_z \zeta \,\partial_{\bar{z}}\zeta
\,-\,\frac{1}{\pi}\;\frac{m\,(\,m-1\,)}{(\,1-z\bar{z}\,)^2}\;\zeta^2
\end{equation}
and the equation for the Green function that follows from it
\begin{equation}
\frac{2}{\pi}\,\partial_z \partial_{\bar{z}} \, g_\zeta(z,z')
\,-\,\frac{2}{\pi}\;\frac{m\,(\,m-1\,)}{(\,1-z\bar{z}\,)^2}\; g_\zeta(z,z')
\;=\;\delta^2(\,z-z'\,)~.
\end{equation}
Such an equation is invariant under $SU(1,1)$.
The explicit form of the Green function $g_\zeta(z,z')$ is
\begin{equation}
 \label{PVCregulator}
g_\zeta(z,z')=-\,\frac{1}{2}\,\frac{\Gamma(m)^2}{\Gamma(2m)}\,
(\,1-\eta\,)^m\,\hspace{-.3cm}\phantom{F}_2 F_1(m,m,2m;1-\eta\,)
\end{equation}
where $\eta(z,z')$ is the invariant (\ref{eta(z,z_n)}) and
$\hspace{-.3cm}\phantom{F}_2 F_1$ is the 
hypergeometric function.\\
Note that, for fixed $\eta \neq 0$, $g_\zeta(z,z') \rightarrow 0$
when $m \rightarrow \infty$. Because of this property, all the
diagrams containing $g_\zeta(z,z')$ with $z \neq z'$ vanish when
$m \rightarrow \infty$.\\
The divergence of $g_\zeta(z,z')$ at coincident points (\,$z \rightarrow
z'$, i.e. $\eta \rightarrow 0$\,) cancels out the one of $g(z,z')$ as
follows
\begin{equation}
\lim_{z' \rightarrow z}\,\left[\,g_(z,z')+g_\zeta(z,z')\,\right]\,=\,
-1+\gamma+\psi(m)
\end{equation}
where $\psi(x)=\Gamma'(x)/ \Gamma(x)$ is the digamma function.\\
Thus the Pauli-Villars regulator (\ref{PVCregulator}) generates a
$C$ regulator, where the constant $C$ diverges as $\log(m)$ 
when $m \rightarrow +\infty$.\\
A different Pauli-Villars regulating field can be introduced
through the lagrangian
\begin{equation}
-\,\frac{1}{\pi}\,\partial_z \zeta \,\partial_{\bar{z}}\zeta
\,-\,\frac{M^2}{4\pi}\;\zeta^2
\end{equation}
that gives the following equation for the Green function
\begin{equation}
\frac{2}{\pi}\,\partial_z \partial_{\bar{z}} \, g_\zeta(z,z')
\,-\,\frac{M^2}{2\pi}\; g_\zeta(z,z')
\;=\;\delta^2(\,z-z'\,)~.
\end{equation}
If we set
\begin{equation}
g_\zeta(z,z')=-K_0(\,2M\,|\,z-z'\,|\,)+R(z,z')
\end{equation}
being $K_0$ the modified Bessel function of the second kind, then $R(z,z')$
satisfies the homogeneous elliptic equation
\begin{equation}
4 \,\partial_z \partial_{\bar{z}} \, R(z,z')
\,-\,M^2\, R(z,z')\;=\;0
\end{equation}
with boundary condition
\begin{equation}
R(z,z')= K_0(\,2M\,|\,z-z'\,|\,) \qquad |z| \rightarrow 1
\end{equation}
that gives $g_\zeta(z,z') \rightarrow 0$ when $|z| \rightarrow 1$.\\
Using the maximum principle for this kind of equations
\cite{CourantHilbertII}, we obtain
\begin{equation}
|\,R(z,z')\,| \;\leqslant 
 \; \mathrm{max}_{\begin{array}{l}
\vspace{-.75cm}~\\
\hspace{-.95cm} \vspace{-.4cm} \scriptscriptstyle |w|=1 \end{array}}\, 
\hspace{-.3cm}  K_0(\,2M\,|\,w-z'\,|\,) 
\end{equation}
for every fixed $z$ and $z'$ inside the disk $\Delta$.\\
Thus, for fixed $z$ and $z'$ inside the unit disk, we have that 
$R(z,z') \rightarrow 0$  when $M \rightarrow +\infty$, 
being $K_0(\,M\,|w-z'|\,) \rightarrow 0$ for $M \rightarrow +\infty$.\\
The field $\zeta$ provides the regulating field and, in this case, we
find
\begin{equation}
\lim_{z' \rightarrow z}\,\left[\,g(z,z')+g_\zeta(z,z')\,\right]\,=\,
\log\,(\,1-z\bar{z}\,)-1+\gamma+\log M+ R(z,z)~.
\end{equation}
This reproduces the ZZ regularization
\cite{ZZ:Pseudosphere} by a proper subtraction of the divergent term
since, for every fixed $z$ inside the disk, $R(z,z)= O(e^{-M})$.\\
In closing this section, we notice that, 
using (\ref{geometric action pseudosphere phiM}), the decomposition of 
$S_{\Delta,N,M}[ \,\phi_M,\phi_B^{cl}\,]$ 
given in (\ref{action phiM N=1}) and the expression
for the one point function (\ref{geometric one-point phiM}), 
one can write the $N$ point vertex functions 
(\ref{geometric N-point function}) as follows
\begin{eqnarray}
 \label{Npointgeomsintetic}
\left\langle \, V_{\alpha_1}(z_1)\dots  V_{\alpha_N}(z_N) \,\right\rangle
=\left[\;\prod_{n=1}^N
\left\langle\,V_{\alpha_n}(z_n)\,\right\rangle \;\right]\,
e^{2 \sum_{n}\alpha_n \sum_{m \neq n} \alpha_m \,g(z_n,z_m)}\,
\frac{\left\langle\,e^{-\widetilde{S}_{\Delta,N,M}}\,
\right\rangle_{z_1,\dots, z_N}}{\prod_{n=1}^N \left\langle\,
e^{-\widetilde{S}_{\Delta,1,M}}\,\right\rangle_{z_n}} \nonumber \\
\end{eqnarray}
where the mean values in the ratio 
are taken with respect to (\ref{standard action}).\\
If we compute formally the limit of the previous expression e.g. for 
$|z_1| \rightarrow 1$, we can verify the cluster decay at large distance
\begin{equation}
\lim_{|z_1| \rightarrow 1} 
\left\langle \,V_{\alpha_1}(z_1)\dots  V_{\alpha_N}(z_N)\,\right\rangle =
\left\langle \,V_{\alpha_1}(z_1)  \,\right\rangle\,
\left\langle \,V_{\alpha_2}(z_2) \dots V_{\alpha_N}(z_N) \,\right\rangle
\end{equation} 
that is the boundary condition used by ZZ \cite{ZZ:Pseudosphere} 
to get the one point function through the bootstrap method.\\

\section{Two point function}
 \label{Two point}

To gain further insight into the relation between the two approaches,
we provide the perturbative computation of the two point function. \\ 
First we compute the complete one loop order and some results to two loop 
within the standard approach for generic $\alpha_1$ and $\alpha_2$. 
Then, we compare these results with the ones
obtained within the geometric approach for generic $\alpha_1$ and $\alpha_2$.\\
A perturbative check of the exact formula conjectured  for the
auxiliary two point function 
$\left\langle\,V_{\alpha}(z') \;V_{-b/2}(z)\,\right\rangle$
\cite{ZZ:Pseudosphere, FZZ, Teschner} will be given at the end of
this section.\\
As pointed out in \cite{ZZ:Pseudosphere}, it is more efficient
to compute the ratio
\begin{equation}
 \label{g_a1a2definition}
g_{\alpha_1,\alpha_2}(\eta) =
\frac{\left\langle\,V_{\alpha_1}(z_1)
  \;V_{\alpha_2}(z_2)\,\right\rangle }{\left\langle
  \,V_{\alpha_1}(z_1) \,\right\rangle \, 
\left\langle \,V_{\alpha_2}(z_2) \,\right\rangle }
\end{equation}
where $\left\langle\,V_{\alpha_1}(z_1) \;V_{\alpha_2}(z_2)\,\right\rangle$ 
represents the full correlator and not only  the connected component.\\
Taking into account once more the cumulant expansion 
\begin{equation}
 \label{cumulant2point}
\log \left[\,\rule{0pt}{.4cm}g_{\alpha_1,\alpha_2}(\eta)\,\right]=
\sum_{k_1,k_2=1}^{\infty}
\frac{(\,2\alpha_1)^{k_1}}{k_1 !}\,\frac{(\,2\alpha_2)^{k_2}}{k_2 !}
\,\,M_{k_1,k_2}(\eta)
\end{equation}
we compute perturbatively $M_{k_1,k_2}(\eta)$.\\
From (\ref{g_a1a2definition}), it can be easily seen that 
the background field does not contribute to
$g_{\alpha_1,\alpha_2}(\eta)$.\\
The graphs contributing to $M_{1,1}$ up to one loop are

\begin{equation}
\begin{array}{lll}
 \label{M11standard}
\vspace{.6cm}
M_{1,1} \hspace{.2cm} & = & \hspace{.4cm}
\;\left[\;\rule{0pt}{.8cm} \hspace{.4cm} 
\begin{minipage}[c]{2.6cm} 
  \includegraphics[width=2cm]{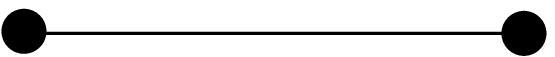}
\end{minipage} 
\right]
  \\ 
\vspace{.4cm}
& + &b^2\;\left[ \hspace{.4cm} \rule{0pt}{1.2cm}
\begin{array}{ccc}
\begin{minipage}[c]{3.5cm} 
  \includegraphics[width=3cm]{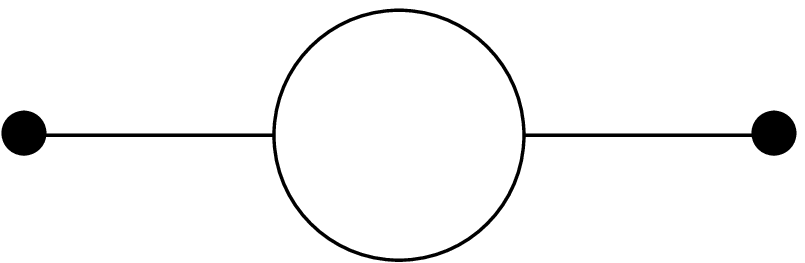}
\end{minipage} 
&
\hspace{.3cm}
\begin{minipage}[c]{3.5cm} 
  \includegraphics[width=3cm]{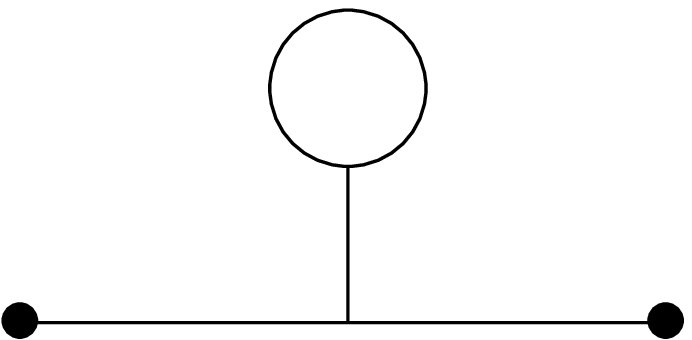}
\end{minipage} 
&
\hspace{.3cm}
\begin{minipage}[c]{4.5cm} 
  \includegraphics[width=3cm]{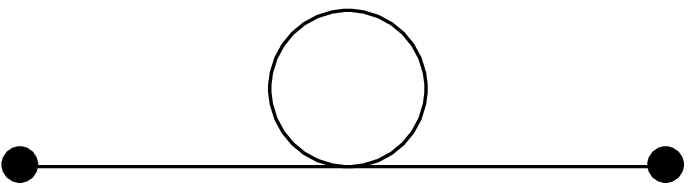}
\end{minipage} 
\end{array}
 \hspace{-.9cm}\right]\\
& + & O(b^4)~.
\end{array}
\end{equation}

\noindent 
As requested by the standard approach \cite{ZZ:Pseudosphere},
we have computed them by using the ZZ regulator, obtaining
\begin{eqnarray}
M_{1,1} &=&
\left\langle\,\chi(z_1) \;\chi(z_2)\,\right\rangle \,-\,
\left\langle\,\chi(z_1)\,\right\rangle\,
\left\langle\,\chi(z_2)\,\right\rangle
\nonumber \\
&=& \rule{0pt}{.8cm}
g(\eta)\,
+b^2\,\Bigg(\,\frac{3}{2}+\frac{\eta^2\,\log^2\eta}{2\,(1-\eta)^2}\,-
\,\frac{1+\eta}{1-\eta}\;\textrm{Li}_2(1-\eta)\,\Bigg)+\,O(b^4)~.
\end{eqnarray}
As expected, the limit of the $O(b^2)$ contribution to 
$M_{1,1}$ for $\eta \rightarrow 0$ gives exactly the order 
$O(b^2)$ of $G_2$.\\
Concerning $M_{2,1}$, its first order expansion is given by the
following graph

\begin{equation}
M_{2,1}\hspace{.3cm}   = \hspace{.3cm}
b\;\left[\; \rule{0pt}{1.2cm}\hspace{.4cm} 
\begin{minipage}[c]{3.5cm} 
  \includegraphics[width=3cm]{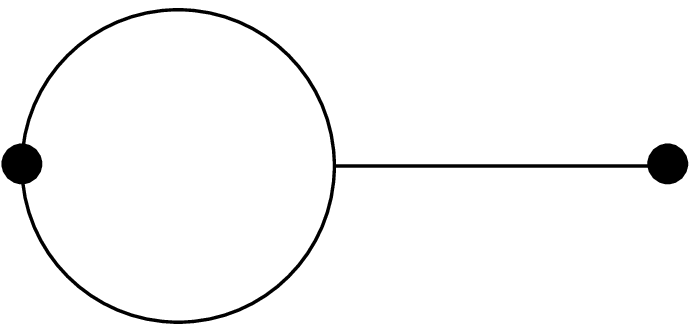}
\end{minipage} 
\right] 
\hspace{.2cm} +\hspace{.2cm}  O(b^3)
\end{equation}

\rule{0pt}{.2cm}

\noindent and it gives \cite{ZZ:Pseudosphere}
\begin{eqnarray}
M_{2,1} &=&  
\langle\,\chi^2(z_1) \;\chi(z_2)\,\rangle \,-\,\langle\,\chi^2(z_1)\,\rangle\,
\langle\,\chi(z_2)\,\rangle\,\nonumber \\
& & \hspace{2.8cm}
-\;2\,\langle\,\chi(z_1) \;\chi(z_2)\,\rangle \,\langle\,\chi(z_1)\,\rangle\,+
2\,\langle\,\chi(z_1)\,\rangle^2\,\langle\,\chi(z_2)\,\rangle
\nonumber \\
&=& \rule{0pt}{1cm}
b\;\Bigg(\,\frac{\eta\,\log^2\eta}{(1-\eta)^2}\,-1\,\Bigg)+\,O(b^3)~.
\end{eqnarray}

\noindent
We have considered also the two loop graphs that give the 
first order of $M_{3,1}$

\begin{equation}
M_{3,1}\hspace{.3cm}   = \hspace{.3cm}
b^2\;\left[\;\rule{0pt}{1.2cm} \hspace{.4cm} 
\begin{minipage}[c]{3.5cm} 
  \includegraphics[width=3cm]{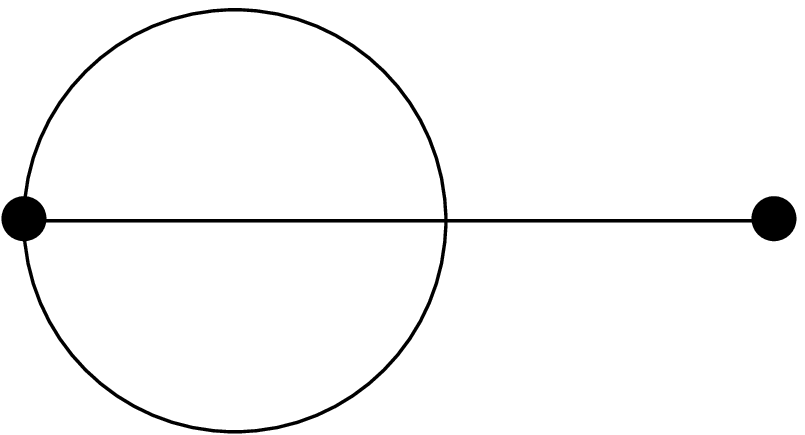}
\end{minipage} 
\hspace{.8cm}
\begin{minipage}[c]{3.5cm} 
  \includegraphics[width=3cm]{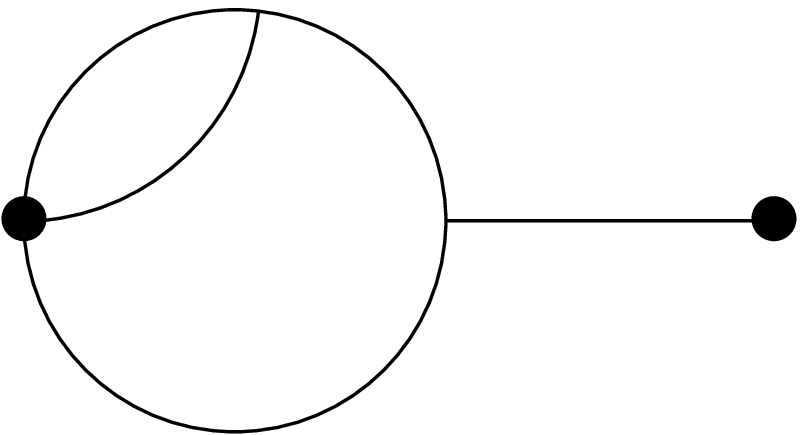}
\end{minipage} 
\right] 
\hspace{.2cm} +\hspace{.2cm}  O(b^4)~.
\end{equation}

\noindent The result is

\begin{equation}
M_{3,1}
=b^2\;\Bigg(-\,\frac{\eta\,(1+\eta)\,\log^3 \eta}{(1-\eta)^3}\,-2\,
\Bigg)+\,O(b^4)~.
\end{equation}

\noindent
To compute the graphs contributing to $O(b^2)$ in $M_{1,1}$ and to 
$O(b^2)$ in $M_{3,1}$, we have employed the same technique developed
in \cite{tetrahedron}, first giving (\,in analogy to the
Gegenbauer method; see e.g. \cite{Kotikov}\,) an harmonic expansion of the 
Green function (\ref{g(z,z_n)}) and then reducing the graphs, through angular
integrations, to radial integrals.\\
Within the geometric approach, by using (\ref{Npointgeomsintetic}) 
with $N=2$, we introduce
$g^\textrm{g}_{\alpha_1,\alpha_2}(\eta)$ as we have done in
(\ref{g_a1a2definition}) for the standard approach, obtaining
\begin{equation}
 \label{g_a1a2geometric}
g^\textrm{g}_{\alpha_1,\alpha_2}(\eta)
\hspace{.1cm}=\hspace{.1cm}
\frac{\left\langle\,V_{\alpha_1}(z_1)
  \;V_{\alpha_2}(z_2)\,\right\rangle }{\left\langle
  \,V_{\alpha_1}(z_1) \,\right\rangle \, 
\left\langle \,V_{\alpha_2}(z_2) \,\right\rangle }
\hspace{.1cm}=\hspace{.1cm}
\,\,e^{4 \, \alpha_1 \alpha_2 \,g(z_1,z_2)}\;\frac{\left\langle\,
  e^{-\widetilde{S}_{\Delta,2,M}}\,\right\rangle_{z_1
    ,\,z_2}}{\left\langle\,e^{-\widetilde{S}_{\Delta,1,M}}\,
\right\rangle_{z_1}\left\langle\,e^{-\widetilde{S}_{\Delta,1,M}}\,
\right\rangle_{z_2}}\,~.   
\end{equation} 
Notice that in this ratio the possible ambiguities 
$\lambda_n$ in the subtraction terms of the geometric action, 
mentioned in Section \ref{One point function}, cancel out.\\ 
Again, we consider the cumulant expansion of 
$g^\textrm{g}_{\alpha_1,\alpha_2}(\eta)$, which defines the functions
$M_{k_1,k_2}^{\textrm{g}}(\eta)$, as done in (\ref{cumulant2point}) 
for the standard approach.\\
From (\ref{g_a1a2geometric}), we can get the functions
$M_{k_1,k_2}^{\textrm{g}}(\eta)$ within the geometric approach
\begin{eqnarray}
M_{1,1}^{\textrm{g}}&=& 
g(\eta)+\,\frac{1}{2^2}\;\Bigg\{\,
-\,\left\langle\, \widetilde{S}_{\Delta,2,M}^{(1,1)}\,\right\rangle 
\,+\,\Big\langle\, \widetilde{S}_{\Delta,2,M}^{(1,0)}\;
\widetilde{S}_{\Delta,2,M}^{(0,1)}\,\Big\rangle\,
\,-\,
\Big\langle\,\widetilde{S}_{\Delta,2,M}^{(1,0)}\,\Big\rangle\,
\Big\langle\,\widetilde{S}_{\Delta,2,M}^{(0,1)}\,\Big\rangle
\,\Bigg\} \nonumber \\
& & \\
M_{2,1}^{\textrm{g}}&=& 
\frac{1}{2^3}\;\Bigg\{\,
-\,\left\langle\, \widetilde{S}_{\Delta,2,M}^{(2,1)}\,\right\rangle 
\,+\,\Big\langle\, \widetilde{S}_{\Delta,2,M}^{(2,0)}\;
\widetilde{S}_{\Delta,2,M}^{(0,1)}\,\Big\rangle
\,+\,2\,\Big\langle\, \widetilde{S}_{\Delta,2,M}^{(1,0)}\;
\widetilde{S}_{\Delta,2,M}^{(1,1)}\,\Big\rangle 
\nonumber \\
& & 
\rule{0pt}{.5cm}
-\,\,\Big\langle\,\left( \widetilde{S}_{\Delta,2,M}^{(1,0)}\right)^2\;
\widetilde{S}_{\Delta,2,M}^{(0,1)}\,\Big\rangle\,-\,
\Big\langle\,\widetilde{S}_{\Delta,2,M}^{(2,0)}\,\Big\rangle\,
\Big\langle\,\widetilde{S}_{\Delta,2,M}^{(0,1)}\,\Big\rangle\,-\,
2\,\Big\langle\,\widetilde{S}_{\Delta,2,M}^{(1,0)}\,\Big\rangle\,
\Big\langle\,\widetilde{S}_{\Delta,2,M}^{(1,1)}\,\Big\rangle
\nonumber \\
& & 
\rule{0pt}{.8cm}
+\,\,
\Big\langle\,\left( \widetilde{S}_{\Delta,2,M}^{(1,0)}\right)^2\Big\rangle\,
\Big\langle\,\widetilde{S}_{\Delta,2,M}^{(0,1)}\,\Big\rangle\,+\,2\,
\Big\langle\,\widetilde{S}_{\Delta,2,M}^{(1,0)}\;
\widetilde{S}_{\Delta,2,M}^{(0,1)}\,\Big\rangle\,
\Big\langle\,\widetilde{S}_{\Delta,2,M}^{(1,0)}\,\Big\rangle\,
\nonumber \\
& & 
\rule{0pt}{.8cm}
-\,2\,
\left(\,\Big\langle\,\widetilde{S}_{\Delta,2,M}^{(1,0)}\,\Big\rangle\,\right)^2
\Big\langle\,\widetilde{S}_{\Delta,2,M}^{(0,1)}\,\Big\rangle\;\Bigg\}
\\
M_{3,1}^{\textrm{g}}&=& \dots \nonumber  
\end{eqnarray}
where
\begin{equation}
\widetilde{S}_{\Delta,2,M}^{(k_1,k_2)}=
\left.\frac{\partial^{k_1}}{\partial \alpha_1^{k_1}}\,
\frac{\partial^{k_2}}{\partial \alpha_2^{k_2}}\,
\widetilde{S}_{\Delta,2,M}\,
\right|_{\alpha_1=\alpha_2=0}~.
\end{equation}
The graphs contributing to $M_{k_1,k_2}^{\textrm{g}}(\eta)$ up to 
$b_\textrm{g}^2$ included are shown below.

\begin{equation}
 \label{M11geomgraphs}
\vspace{-.3cm}
\begin{array}{lll}
\vspace{.6cm}
M_{1,1}^{\textrm{g}} \hspace{.2cm} & = & \hspace{.4cm}
\;\left[\;\rule{0pt}{.8cm} \hspace{.4cm} 
\begin{minipage}[c]{2.6cm} 
  \includegraphics[width=2cm]{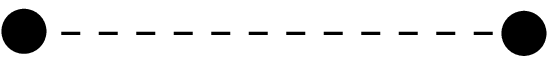}
\end{minipage} 
\right]
  \\ 
\vspace{.4cm}
& + & b_\textrm{g}^2 \;\left[ \hspace{.4cm}\rule{0pt}{1.2cm}
\begin{array}{ccc}
\begin{minipage}[c]{3.5cm} 
  \includegraphics[width=3cm]{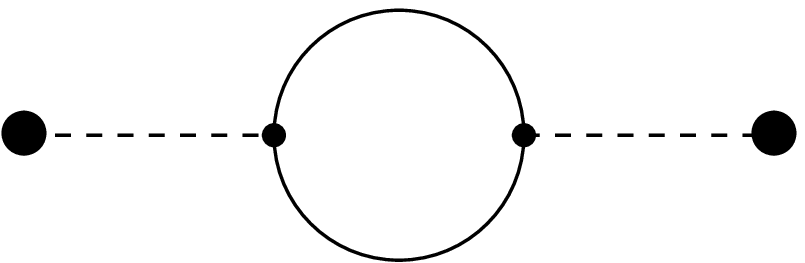}
\end{minipage} 
&
\hspace{.3cm}
\begin{minipage}[c]{3.5cm} 
  \includegraphics[width=3cm]{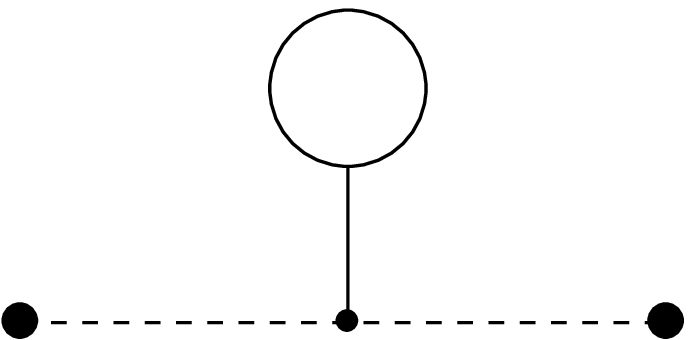}
\end{minipage} 
&
\hspace{.3cm}
\begin{minipage}[c]{4.5cm} 
  \includegraphics[width=3cm]{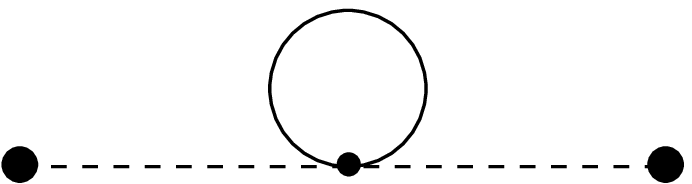}
\end{minipage} 
\end{array}
 \hspace{-.9cm}\right] \\
& + & O(b_\textrm{g}^4)~.
\end{array}
\end{equation}
\begin{equation}
\rule{0pt}{1.7cm}
\hspace{-6.7cm}
M_{2,1}^{\textrm{g}}\hspace{.45cm}   = \hspace{.3cm}
b_\textrm{g}\;\left[\; \rule{0pt}{1.2cm}\hspace{.4cm} 
\begin{minipage}[c]{3.5cm} 
  \includegraphics[width=3cm]{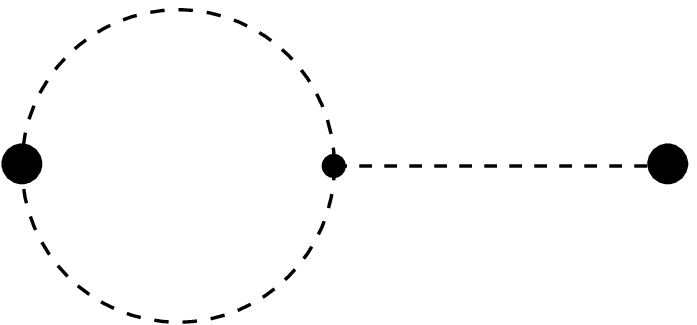}
\end{minipage} 
\right] 
\hspace{.2cm} +\hspace{.2cm}  O(b_\textrm{g}^3)
\end{equation}
\begin{equation}
\rule{0pt}{1.7cm}
\hspace{-2.15cm}
M_{3,1}^\textrm{g}\hspace{.45cm}   = \hspace{.3cm}
b_\textrm{g}^2\;\left[\;\rule{0pt}{1.2cm} \hspace{.4cm} 
\begin{minipage}[c]{3.5cm} 
  \includegraphics[width=3cm]{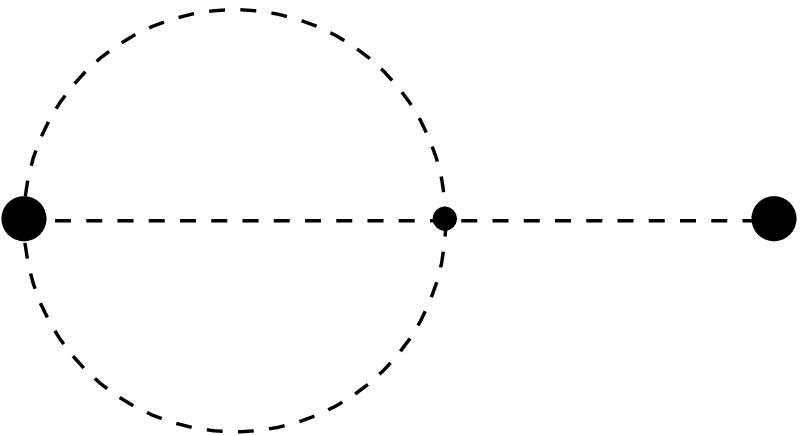}
\end{minipage} 
\hspace{.8cm}
\begin{minipage}[c]{3.5cm} 
  \includegraphics[width=3cm]{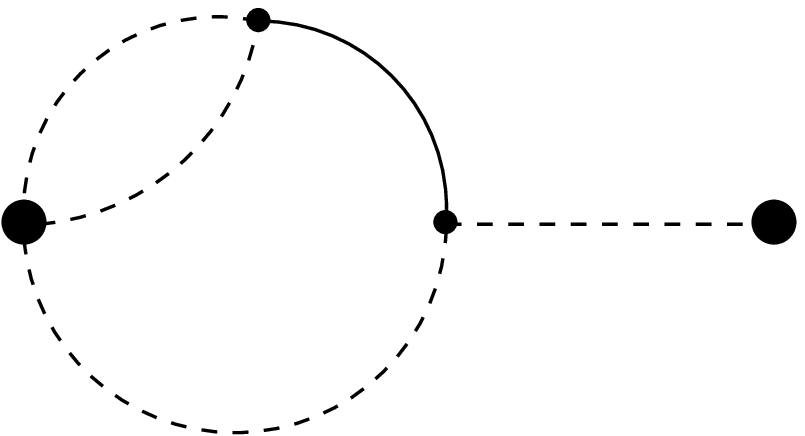}
\end{minipage} 
\right] 
\hspace{.2cm} +\hspace{.2cm}  O(b_\textrm{g}^4)~.
\end{equation}

\noindent \rule{0pt}{.5cm}\\
Comparing $M_{k_1,k_2}^{\textrm{g}}(\eta)$ with the functions 
$M_{k_1,k_2}(\eta)$ obtained within the standard approach,
we have found no differences for $M_{2,1}^{\textrm{g}}(\eta)$ 
and $M_{3,1}^{\textrm{g}}(\eta)$ up to $O(b^2)$ included, while a discrepancy 
with the standard approach arises comparing
$M_{1,1}^{\textrm{g}}(\eta)$ with $M_{1,1}(\eta)$. 
This difference is due to the fact that, using the Hadamard regularization,
the second and the third graph of the $O(b_\textrm{g}^2)$ contribution to
$M_{1,1}^{\textrm{g}}(\eta)$ given in (\ref{M11geomgraphs}) sum up to zero,
while, on the other hand, within the standard approach with ZZ regulator,
the second and the third graph in the $O(b^2)$ contribution to
$M_{1,1}(\eta)$ shown in (\ref{M11standard}) provide a non trivial function.\\
We have, up to order $b^2$
\begin{eqnarray}
M_{1,1}^{\textrm{g}}(\eta) &=&
M_{1,1}(\eta) 
+b^2 \,\frac{2}{3}\,\bigg(-1+\log(1-\eta)
+\,\frac{\eta \,\log\,\eta}{1-\eta}\,  
+\,\frac{(1+\eta)\,\log(1-\eta)\,\log\,\eta}{2\,(1-\eta)}\nonumber \\
& & \hspace{3.83cm}
+\,\frac{1+\eta}{1-\eta}\;\textrm{Li}_2(1-\eta)\,\bigg)+O(b^4)~.
\end{eqnarray}

\noindent It is interesting to compare such results with the exact formula 
conjectured in \cite{ZZ:Pseudosphere,FZZ,Teschner} for 
$G_{\alpha,-b/2}(z',z)=\left\langle\,V_{\alpha}(z')\;V_{-b/2}(z)\,
\right\rangle$, by setting $\alpha_1=\alpha$,
$\alpha_2=-b/2$, $z_1=z'$ and $z_2=z$.\\
The conjectured formula can be written equivalently in the upper half plane 
as
\cite{ZZ:Pseudosphere}
\begin{equation}
 \label{twopointconjectureUHP}
G_{\alpha,-b/2}(\xi',\xi)=\frac{\left|\,\xi'-\bar{\xi}'\,
\right|^{2\Delta_{-b/2}-2\Delta_\alpha}}{\left|\,\xi-\bar{\xi}'\,
\right|^{4\Delta_{-b/2}}}\;\,
U(\alpha)\,U(-b/2)\;(\,1-\eta\,)^{1+3\,b^2/2}\,g_{\alpha,-b/2}(\eta)
\end{equation}
and in the unit disk $\Delta$ representation as
\begin{equation}
 \label{twopointconjectureDelta}
G_{\alpha,-b/2}(z',z)=\frac{\left(\,1-z'\bar{z}'\,\right)^{2\Delta_{-b/2}-
2\Delta_\alpha}}{\left(\,1-z\bar{z}'\,\right)^{4\Delta_{-b/2}}}\;\,
U(\alpha)\,U(-b/2)\;(\,1-\eta\,)^{1+3\,b^2/2}\,g_{\alpha,-b/2}(\eta)
\end{equation}
where $U(\alpha)$ is the structure constant of the one
point function conjectured in \cite{ZZ:Pseudosphere}.\\
Notice that, to get (\ref{twopointconjectureUHP}) 
or equivalently (\ref{twopointconjectureDelta}),
the ZZ boundary conditions (\,i.e. the cluster decay at large distance 
$B^{(-)}(\alpha)=U(\alpha)\,U(-b/2)\,$)
have been used \cite{ZZ:Pseudosphere}.\\
From
\begin{equation}
G_{\alpha,-b/2}(z',z)=
\big\langle \,V_{\alpha}(z') \,\big\rangle \;
\big\langle \,V_{-b/2}(z) \,\big\rangle \, 
\;g_{\alpha,-b/2}(\eta)
\end{equation}
where
\begin{equation}
g_{\alpha,-b/2}(\eta)
=\eta^{\alpha b}\,_2F_1(\,1+b^2,2\alpha b,2+2b^2;1-\eta\,)
\end{equation}

\noindent the cumulant expansion of $g_{\alpha,-b/2}(\eta)$ in
$\alpha$ up to $O(b^3)$ is
\begin{eqnarray}
\log \left[\,\rule{0pt}{.4cm}g_{\alpha,-b/2}(\eta)\,\right] & = & 
\nonumber \\
& \hspace{-2cm}= & 
\hspace{-1.1cm}
\rule{0pt}{.8cm}\alpha\,\left[\,b \,\Big(-2\,g(\eta)\Big)
+b^3\,\Bigg(-4+\frac{\eta\,\log^2\eta}{1-\eta}\,+2\,\frac{1+\eta}{1-\eta}\;
\textrm{Li}_2(1-\eta)\Bigg)+\,O(b^5)\,\,\right] \nonumber \\
& & 
\hspace{-1.1cm}
\rule{0pt}{.8cm} +\,\alpha^2 \,
\left[\,b^2\,\Bigg(\,2-2\,\frac{\eta\,\log^2\eta}{(1-\eta)^2}\,\Bigg)+
\,O(b^4)\,\,\right]
\nonumber \\
& & 
\hspace{-1.1cm}
\rule{0pt}{.8cm} +\,\alpha^3 \,
\left[\,b^3\,\Bigg(\,\frac{8}{3}+\frac{4}{3}\,\frac{\eta\,(1+\eta)\,
\log^3\eta}{(1-\eta)^3}\,\Bigg)+\,O(b^5)\,\,\right]\,+\,O(\alpha^4)~. 
\end{eqnarray} 
It agrees perfectly with our perturbative results obtained within the 
standard approach and, as discussed above, it disagrees with the 
ones found within the geometric approach.\\

\section{Invariance under background transformations}

In order to improve our understanding of the two approaches,
we shall examine in this section the dependence of the results
on the choice of the background. We shall find that both
the approaches are background independent.\\
The perturbative quantization of the lagrangian (\ref{geometric action
  pseudosphere phiM}) with the boundary conditions of the pseudosphere
has been carried through starting from the classical background solution
$\phi_B^{cl}$ and perturbatively expanding around it. 
This has been done both within the
standard approach \cite{ZZ:Pseudosphere, tetrahedron} and  within
the geometric approach.\\
In this section, we shall examine the independence of the results on
the choice of the background. We shall provide this calculation both at
the formal level, by using the equation of motion for $\phi_M$, and at
the perturbative level, i.e. expanding the results perturbatively in
the background field around $\phi_B^{cl}$ and verifying that these
results are independent on such variations. This happens both within the
standard approach and within the geometric approach, independently on
the choice of the regulators. 
We have obtained the  same results by employing the related Pauli-Villars 
regulator fields, described in Section \ref{One point function}.\\
First we give a formal proof of the background invariance
within the standard approach. Using the background field 
method, i.e. decomposing the Liouville field $\phi$ as the sum of a 
background field $\phi_B$ and a quantum correction $\chi$, the action on the
pseudosphere with a generic background becomes \cite{ZZ:Pseudosphere}
\begin{equation}
S[ \,\phi\,]=S_B[ \,\phi_B\,]+S_\chi[\,\chi,\phi_B\,]
\end{equation}
where
\begin{equation}
S_B[ \,\phi_B\,]=
\int_{\Delta} \left[
\,\frac{1}{\pi}\, \partial_z \phi_B \,\partial_{\bar{z}}\phi_B+\mu
e^{2b\phi_B}\,\right]\, d ^2 z 
\end{equation}
and
\begin{equation}
S_\chi[\,\chi,\phi_B\,]=
 \int_{\Delta}  \left[\,\frac{1}{\pi}\,
  \partial_z  \chi \,\partial_{\bar{z}}\chi +
  \mu e^{2b \phi_B}\Big(e^{2b \chi}-1 \Big)-
  \frac{2}{\pi}\,\chi  \, \partial_z \partial_{\bar{z}} \phi_B
\,\right]\,d ^2z~.  
\end{equation}
Within the standard approach the correlation functions 
of the vertex operators are given by
\begin{equation}
 \label{standard correlation functions}
\left\langle \, V_{\alpha_1}(z_1) \dots V_{\alpha_N}(z_N)\,\right\rangle
 \,=\,\frac{\displaystyle \int_{\mathcal{C}(\Delta)} \hspace{-.3cm}
\mathcal{D}\,[\, \chi \,]\,\, 
e^{-S_\chi  \left[ \,\chi,\phi_B \,\right]\,+
\sum_n^N 2\alpha_n \left(\,\chi(z_n)+\phi_B(z_n)\,\right)}}{\displaystyle 
\int_{\mathcal{C}(\Delta)}  \hspace{-.3cm} \mathcal{D}\,[\, \chi \,]\,\, 
e^{- S_\chi  \left[ \,\chi,\phi_B \,\right]}}\, 
=\, \frac{Z[\,J\,]}{Z[\,0\,]}
\end{equation}
because the background contribution simplifies.\\
Under a variation $\delta \phi_B$ with compact support of the generic
background field $\phi_B$, we have
\begin{equation}
 \label{firstvariation}
\delta \,
\left\langle \, V_{\alpha_1}(z_1) \dots V_{\alpha_N}(z_N)\,\right\rangle\,=\,
\frac{\delta\,Z[\,J\,]}{Z[\,0\,]}\,-
\,\frac{\delta\,Z[\,0\,]}{Z[\,0\,]}\,
\frac{Z[\,J\,]}{Z[\,0\,]}~.
\end{equation}
From (\ref{standard correlation functions}), we get 
\begin{eqnarray}
 \label{dZ[J]}
\hspace{0cm}
\delta\,Z[\,J\,] \hspace{-.3cm} \,& = & \, \hspace{-.3cm}
\int_{\mathcal{C}(\Delta)} \hspace{-.3cm}
\mathcal{D}\,[\, \chi \,]\, \int_{\Delta}  
\left[\, \rule{0pt}{.8cm} \frac{2}{\pi}\,  \partial_z \partial_{\bar{z}} \chi- 
  2 b \,\mu e^{2b \phi_B}\Big(e^{2b\chi}-1 \Big) \right. \nonumber \\
 & & \hspace{3cm}
\left. + 2 \sum_{n=1}^N \alpha_n \delta^2(z-z_n)\,\right]\,\delta \phi_B\,d ^2z
\; e^{-S_\chi  \left[ \,\chi,\phi_B \,\right]\,+
\sum_n^N 2\alpha_n(\,\chi(z_n)+\phi_B(z_n)\,)} \nonumber \\
& & 
\end{eqnarray} 
that, due to the equation of motion in presence of sources 
for the field $\chi$, becomes 
\begin{equation}
\delta\,Z[\,J\,]\; = \;Z[\,J\,]\int_{\Delta}  \left[\,-\,\frac{2}{\pi}\,  
\partial_z \partial_{\bar{z}} \phi_B + 
  2 b \,\mu e^{2b \phi_B}\,\right]\delta \phi_B\,d ^2z 
\end{equation}
i.e. the variation of $Z[\,J\,]$ is given by a numerical factor that
multiplies
$Z[\,J\,]$.
The same structure and the same numerical factor appear 
for the variation of $Z[\,0\,]$.\\
Thus, the first variation (\ref{firstvariation})
of the correlation functions (\ref{standard correlation functions}) 
under changes of the background  vanishes. Since such
a variation is null starting from a generic background field, we
have the independence of the correlation functions from $\phi_B$.\\
Obviously, the above general reasoning is only formal since the theory
contains divergencies.\\
On the other hand, ZZ \cite{ZZ:Pseudosphere} proved the validity of
the equations of motion on the background $\phi_B^{cl}$ up to order
$b^3$ using their particular regulator and suggested their validity to
all orders in the perturbation theory. Moreover
starting from $\phi_B^{cl}$, one can explicitly verify that $G_1$ and
$G_2$ do not change under a variation $\delta \phi_B$ with compact
support, to the first order in $\delta \phi_B$. We did it up to
$O(b^3)$ included for $G_1$ and up to $O(b^2)$ included for $G_2$.\\
A completely similar reasoning works for the geometric 
approach.\\
The correlation functions of the Liouville vertex operators are given
by
\begin{equation}
 \label{geometric correlation functions}
\left\langle \, V_{\alpha_1}(z_1) \dots V_{\alpha_N}(z_N)\,\right\rangle
 \,=\,\frac{\displaystyle \int_{\mathcal{C}(\Delta)} \hspace{-.3cm}
\mathcal{D}\,[\, \phi_M \,]\,\, 
e^{-S_{\Delta,N}  \left[ \,\phi\,\right]}}{\displaystyle 
\int_{\mathcal{C}(\Delta)}  \hspace{-.3cm} \mathcal{D}\,[\, \phi_M \,]\,\, 
e^{- S_{\Delta,0}  \left[ \,\phi\,\right]}}\, 
=\, \frac{Z_\textrm{g}[\,J\,]}{Z_\textrm{g}[\,0\,]}
\end{equation}
where $S_{\Delta,N} \left[ \,\phi\,\right]$ is the
geometric action (\ref{geometric action pseudosphere phiM})
with a generic background satisfying the boundary conditions
(\ref{phiB pseudosphere at infty}).  Varying this action under a
variation $\delta \phi_B$ of compact support of the background field,
we get the variation of $Z_\textrm{g}[\,J\,]$
\begin{eqnarray}
 \label{dZ[J] geometric}
\hspace{0cm}
\delta\,Z_\textrm{g}[\,J\,]  \,& = & \, 
\int_{\mathcal{C}(\Delta)} \hspace{-.3cm}
\mathcal{D}\,[\, \phi_M \,]\, \int_{\Delta}  \left[\,  \frac{2}{\pi}\,  
\partial_z \partial_{\bar{z}} \,(\,\phi_M+g_0\,) \right. \nonumber \\
& & \hspace{3cm}
\left.\,-2 b_\textrm{g} \,\mu_\textrm{g} 
e^{2b_\textrm{g} \phi_B}\Big(e^{2b_\textrm{g}\phi_M}-1\Big)\,\right]\,
\delta \phi_B\,d ^2z
\, e^{-S_{\Delta,N}  \left[ \,\phi\,\right]}\nonumber \\
& & +\,Z_\textrm{g}[\,J\,]\,
\left(\,2\sum_{n=1}^N \alpha_n \delta \phi_B(z_n)\,\right)  
\end{eqnarray} 
that, through the equation of motion for $\phi_M$, becomes
\begin{equation}
\delta\,Z_\textrm{g}[\,J\,]\;=
\;Z_\textrm{g}[\,J\,]\int_{\Delta}  \left[\,-\, \frac{2}{\pi}\,  
\partial_z \partial_{\bar{z}} \phi_B + 
  2 b_\textrm{g} \mu_\textrm{g} e^{2b_\textrm{g} \,\phi_B}\,\right]
\delta \phi_B\,d ^2z~.
\end{equation}
Again, taking into account the variation of $Z_\textrm{g}[\,0\,]$, we have
 to the first order in $\delta \phi_B$
\begin{equation}
\delta \,
\left\langle \, V_{\alpha_1}(z_1) \dots V_{\alpha_N}(z_N)\,\right\rangle\,=\,
\frac{\delta\,Z_\textrm{g}[\,J\,]}{Z_\textrm{g}[\,0\,]}\,-
\,\frac{\delta\,Z_\textrm{g}[\,0\,]}{Z_\textrm{g}[\,0\,]}\,
\frac{Z_\textrm{g}[\,J\,]}{Z_\textrm{g}[\,0\,]}\;=\;0
\end{equation}
also in the geometric approach.\\
Within the regulated theory with the invariant regulator $C$ in presence
of the classical background $\phi_B^{cl}$, we have verified the
validity of the equations of motion up to $O(b_\textrm{g}^4)$ included. 
Moreover one can verify that $G_1^\textrm{g}$ and $G_2^\textrm{g}$ 
do not vary respectively up to $O(b_\textrm{g}^4)$ and $O(b_\textrm{g}^3)$ 
included.\\

\section{Geometric approach with ZZ regulator}
 \label{GeometricZZreg}

From the results obtained in the previous sections, one could get 
the impression that the origin of the differences between the 
standard and geometric approach consists  
in the way used to introduce the sources.\\
An important point, already noticed in \cite{Takhtajan:Equivalence} 
and remarked in Section \ref{One point function}, is that,
the Hadamard regularization within the geometric approach provides 
for the quantum conformal dimensions of the cosmological term
$(\,1-b_\textrm{g}^2,1-b_\textrm{g}^2\,)$.
The only way to find quantum dimensions $(\,1,1\,)$ for the 
cosmological term is to adopt also within the geometric approach
the ZZ regulator \cite{ZZ:Pseudosphere}.
With such a regulator, one obtains the same results
of the standard approach to all orders. 
Indeed, there is a one to one
correspondence between the graphs of the two approaches,
except for the $\alpha^2$ contribution to the quantum 
conformal dimensions, 
which is provided by a counterterm of the action within the 
geometric approach and by the one loop graph regularized 
through the ZZ regulator within the standard approach.
Adopting the ZZ regulator within both approaches,
the two treatments are identified for $b=b_\textrm{g}$ and
$\mu=\mu_\textrm{g}$, and consequently
\begin{equation}
\left[\,\pi b_\textrm{g}^2 \mu_\textrm{g} \,\right]^{-\alpha_1/b_\textrm{g}}
e^{-2\alpha_1^2}\,
\left\langle\,e^{-\widetilde{S}_{\Delta,1,M}}\,\right\rangle_{z_1}
=\;
\frac{U(\alpha_1)}{(\,1-z_1\bar{z}_1\,)^{\,2 b \alpha_1}}
= \left[\,\pi b^2 \mu \,\right]^{-\alpha_1/b}
\frac{\left\langle\,e^{2 \alpha_1 \chi (z_1)}\,\right\rangle}{
(\,1-z_1\bar{z}_1\,)^{\,2\alpha_1^2}}
\end{equation}
where the mean value $\left\langle\,\dots\,\right\rangle_{z_1}$ 
is taken with respect to (\ref{standard action}). 
The one point structure constant $U(\alpha_1)$
is given by the exact formula conjectured in \cite{ZZ:Pseudosphere} 
through the application of the bootstrap method.\\
Another difference between the two approaches is the asymptotic behaviour 
of $\left\langle \,\phi(z_1)\,\right\rangle$, defined as
\begin{equation}
\left\langle \, \phi(z_1)\,\right\rangle=\,\frac{1}{2}\,\left.
\frac{\partial}{\partial \alpha_1}
\left\langle\,e^{2\alpha_1\phi(z_1)}\,\right\rangle\,\right|_{\alpha_1=0}~.
\end{equation}
Within the geometric approach with the $C$ regulator  one obtains
\begin{equation}
\left\langle \, \phi(z_1)\,\right\rangle \simeq 
-\,\frac{1}{2b_\textrm{g}}\,\log\,(\,1-z_1\bar{z}_1\,)^2+ \textrm{const}
\end{equation}
that is exactly the boundary condition (\ref{phi pseudosphere at
  infty}) and (\ref{phiB pseudosphere at infty}) with $Q=1/b_\textrm{g}$,
imposed respectively on the Liouville field $\phi$ and on the background field
$\phi_B$.\\
This does not happen within the standard approach  with the 
ZZ regulator \cite{ZZ:Pseudosphere}, where
\begin{equation}
 \label{standard phi asymptotics}
\left\langle \, \phi(z_1)\,\right\rangle \simeq 
-\,\frac{1}{2}\,\left(\,\frac{1}{b}+b\,\right)\,
\log\,(\,1-z_1\bar{z}_1\,)^2+ \textrm{const}~.
\end{equation}
The asymptotics (\ref{standard phi asymptotics}) reproduces the 
boundary behaviour of the background field 
\begin{equation}
\phi_{cl} (z)= -\,\frac{1}{2b}\,\log\,
\left[ \, \pi b^2 \mu \,  (\,1-z\bar{z} \,)^2 \, \right] 
\end{equation}
only qualitatively, i.e. with a different constant 
in front of the logarithm.\\
Thus, the main difference between the standard and the geometric approach
is not so much in the way one uses to introduce the sources, but in the 
regulator chosen to make the divergent graphs finite, which reflects
in itself the quantum nature of the cosmological term,
deeply related the boundary behaviour of 
$\left\langle\,\phi(z_1)\,\right\rangle$.\\

\section*{Conclusions}

In this paper we have compared the standard and 
the geometric approaches to quantum Liouville theory on the sphere
and on the pseudosphere.
Detailed perturbative calculations up to three loops have been 
performed on the pseudosphere in both approaches.

The geometric approach, even if invariant under $SU(1,1)$ group like the
one of \cite{DFJ}, gives different results with respect to 
standard approach \cite{ZZ:Pseudosphere}. A feature of the geometric
approach, already noticed in \cite{Takhtajan:Equivalence},
is that it provides the cosmological term with quantum conformal dimensions
$(\,1-b_\textrm{g}^2,1-b_\textrm{g}^2\,)$, while the standard approach 
gives $(\,1,1\,)$ for them.
Moreover, while the perturbative calculations in the standard approach
agree with the conformal bootstrap formulas of
\cite{ZZ:Pseudosphere,FZZ,Teschner}, 
this does 
not happen for the results obtained within the geometric approach.
Both theories exhibit background invariance, yet they are definitely
different.

A deeper analysis shows that the real
difference between the two 
approaches does not concern the general setting, i.e. geometric action vs. 
standard sources approach, but it lies
in the process of regularization.
Indeed, as remarked in Section \ref{GeometricZZreg}, adopting
the ZZ regulator 
in the perturbative expansion of the geometric action, we have obtained the 
same results of the standard approach, even if the $\alpha^2$ 
contributions to the quantum conformal dimensions have different formal 
origins within the two approaches.

\section*{Acknowledgments}

\noindent We want to thank Damiano Anselmi, Luigi Cantini and 
Domenico Seminara for useful discussions.

\appendix

\section*{Appendix}

  In this appendix, for sake of completeness,
  we shall give the transformation of the geometric action for the
  sphere \cite{Takhtajan:Topics, Takhtajan:Equivalence} 
  into such a form that no limit process appears \cite{CMS:Liouville} 
  and  therefore suitable to perform perturbative computations. 
  This form can be compared with the one derived for the 
  pseudosphere in Section \ref{geometric action on pseudosphere}.\\
  Again we write the Liouville field $\phi$ as a sum of a
  quantum field $\phi_{M}$, a background field $\phi_{B}$ and a
  source field $\phi_{0}$.  Here we shall leave $\phi_{B}$ generic,
  except for the boundary conditions, and $\phi_{0}$ shall be chosen as
  a solution of a linear equation with point sources.\\
  The geometric action in presence of sources at the points
 $z_{1},...,z_{N}$ can be written, using the notation of \cite{ZZ:Sphere},
 as follows
\begin{eqnarray} 
  \label{geometric action Sphere}
\hspace{-1cm}S_{\textrm{P}^1,\,N}[ \,\phi\,] & = & \hspace{-.2cm}
 \lim_{\begin{array}{l}
\vspace{-.9cm}~\\
\hspace{.14cm} \vspace{-.4cm} \scriptscriptstyle \varepsilon \rightarrow 0 \\
\hspace{.2cm} \scriptscriptstyle \!\!R \rightarrow \infty \end{array}}\,
\hspace{-.2cm}
\Bigg\{\int_{\;\Gamma_{R,\varepsilon}} \left[
\,\frac{1}{\pi}\, \partial_z \phi \,\partial_{\bar{z}}\phi+\mu_\textrm{g}
e^{2b_\textrm{g}\phi}\,\right]\, d ^2 z \, \nonumber \\
  &  & \hspace{1.4cm} +\,\frac{Q}{2\pi i} \oint_{\partial\Gamma_R} \phi
\left( \, \frac{d z}{z}- \frac{d \bar{z}}{\bar{z}}\,
\right)+Q^2 \log R^2 \nonumber \\
  &  & \hspace{1.4cm}
-\,\frac{1}{2\pi i}\sum_{n=1}^N \alpha_n\oint_{\partial\gamma_n} \phi
\left( \, \frac{d z}{z-z_n}- \frac{d \bar{z}}{\bar{z}-\bar{z}_n}\,
\right)- \sum_{n=1}^N \alpha_n^2 \log \varepsilon_n^2 \;\Bigg\}\nonumber\\
  &  &  
\end{eqnarray}
where $d^2 z= \left(i d z \wedge d \bar{z}\right)/2$ and 
the domains of integration are $\Gamma_R=\{ |z| \leqslant
R \}$, $\gamma_n=\{ |z-z_n| \leqslant \epsilon_n \}$ and
$\Gamma_{R,\varepsilon}= \Gamma_R \backslash \bigcup_n \gamma_n$.\\
The field $\phi$ is assumed regular at $\infty$ and to transform as
\cite{ZZ:Sphere}
\begin{eqnarray}
 \label{Phi prime}
\phi(z, \bar{z}) & \rightarrow & \phi ' (w,\bar{w})= \phi(z,\bar{z})-
\frac{Q}{2}\log \left| \frac{d w}{d z}\right|^{2}
\end{eqnarray}
under holomorphic coordinate transformations $z \rightarrow w(z)$,
which implies
\begin{eqnarray}
 \label{phi sphere at infty}
\phi(z,\bar{z}) \simeq -\,Q\,\log\,(z\bar{z})+ O(1) 
&\hspace{.5cm} \textrm{for}\hspace{.5cm} & |z| \rightarrow \infty ~.
\end{eqnarray}
The asymptotic behaviour of $\phi$ near the sources is
\begin{eqnarray}
 \label{phi sphere at infty}
\phi(z,\bar{z}) \simeq -\,\alpha_n\,\log\,|z-z_n|^2+ O(1) 
& \hspace{.5cm} \textrm{for}\hspace{.5cm} & z \rightarrow z_n
\end{eqnarray}
as for the pseudosphere.\\
The geometric action in \cite{Takhtajan:Equivalence} is obtained 
for $Q=1/b_\textrm{g}$, but, to perform a comparison with the standard
approach 
\cite{ZZ:Sphere}, we shall keep $Q$ generic for a while.\\
From (\ref{geometric action Sphere}), one can extract the $\mu_\textrm{g}$
dependence of the correlation functions on the sphere by a proper
constant shift of the Liouville field $\phi$, as done in 
\cite{ZZ:Sphere, GoulianLi, Teschner}.
If we define the correlation functions on the sphere not
dividing by $Z_\textrm{g}[\,0\,]$, as suggested in \cite{ZZ:Sphere},
we obtain for such a dependence the following factor
\begin{equation}
\mu_\textrm{g}^{(\,Q-\sum \alpha_n)/b_\textrm{g}}~.
\end{equation}
The equation of motion derived from the geometric action
(\ref{geometric action Sphere}) is the Liouville equation in presence
of sources
\begin{equation}
 \label{Liouville eq. sources sphere}
\partial_z \partial_{\bar{z}}\phi = \pi b_\textrm{g} \,\mu_\textrm{g} 
e^{2b_\textrm{g}\phi}
-\pi \sum_{n=1}^N \alpha_n \delta^2(\,z-z_n\,)~.
\end{equation}
We write \cite{CMS:Liouville}
\begin{equation}
 \label{phi sphere decomposition}
\phi= \phi_M + \phi_0 + \phi_B
\end{equation}
where $\phi_B$ is the background field, regular on the whole plane
\begin{equation}
 \label{phiB sphere at infty}
\phi_B = -\,Q\,\log \,(\,z\bar{z}\,)+c_B+O\left(\,\frac{1}{|\,z\,|}\,\right) 
\qquad \hspace{1cm}
|z| \rightarrow \infty
\end{equation}
and the source field $\phi_0$ is
\begin{equation}
 \label{phi0 sphere}
\phi_0=-\sum_{n=1}^N \alpha_n \, \log \,|\,z-z_n\,|^2-\hat\alpha \,
\phi_B + c_0 
\end{equation}
i.e. the solution of
\begin{equation}
 \label{phi0 sphere eq.}
\partial_z \partial_{\bar{z}}\phi_0 = -\,\pi \sum_{n=1}^N \alpha_n
\delta^2(\,z-z_n\,) -\hat\alpha \, \partial_z \partial_{\bar{z}}\phi_B 
\end{equation}
with $\hat\alpha$ given by
\begin{equation}
 \label{alpha}
 \hat\alpha = \frac{1}{\displaystyle Q}\displaystyle\, \sum_{n=1}^N \alpha_n
\end{equation}
to have $\phi_0$ going to a constant at $\infty$. As a
consequence, also $\phi_M$ goes to a constant at $\infty$.\\
The equation of motion for $\phi_M$ is
\begin{equation}
 \label{phiM sphere eq.}
\partial_z \partial_{\bar{z}}\phi_M  =  \pi b_\textrm{g} \,\mu_\textrm{g} 
e^{2b_\textrm{g}\phi}+
(\,\hat\alpha-1\,)\, \partial_z \partial_{\bar{z}}\phi_B~.
\end{equation}
Performing a number of integrations by part, the geometric action
(\ref{geometric action Sphere}) can be rewritten as
\begin{eqnarray} 
  \label{geometric action Sphere phiM}
\hspace{-.1cm}
S_{\textrm{P}^1,\,N}[ \,\phi\,] \hspace{-.2cm}
& = \hspace{-.2cm}& S_{\textrm{P}^1,B}[ \,\phi_B\,] +
S_{\textrm{P}^1,N,M}[ \,\phi_M,\phi_B\,]  
- \, \frac{ 2-\hat\alpha}{\pi} \,\int_{\textrm{P}^1} \phi_0 \,  
\partial_z \partial_{\bar{z}} \phi_B \, d ^2z
  \nonumber\\
 & & +\sum_n^N \alpha_n \sum_{m \neq n}^N \alpha_m \,\log \, |\,z_n-z_m \,|^2 
     -\sum_n^N \alpha_n \Bigl(c_0-\hat\alpha \,\phi_B(z_n)\Bigr)
     -\,2\sum_n^N \alpha_n\,\phi_B(z_n) \nonumber \\
& &
\end{eqnarray}
where $S_{\textrm{P}^1,B} [ \,\phi_B\,]$ is the
background action \cite{ZZ:Sphere}
\begin{eqnarray}
 \label{standard action Sphere phiB}
S_{\textrm{P}^1,B}[ \,\phi_B\,] & = & \lim_{R \rightarrow \infty}
\Bigg\{\,\int_{\Gamma_R}  
\left[\,\frac{1}{\pi}\, \partial_z \phi \,\partial_{\bar{z}}\phi_B+
\mu_\textrm{g} e^{2b_\textrm{g}\phi_B}\,\right]\, d ^2 z \, \nonumber \\
& & \hspace{1.4cm}
+\,\frac{Q}{2\pi i} \oint_{\partial\Gamma_R} \phi_B
\left( \, \frac{d z}{z}- \frac{d \bar{z}}{\bar{z}}\,
\right)+Q^2 \log R^2\; \Bigg\}~ 
\end{eqnarray}
and $S_{\textrm{P}^1,N,M}[ \,\phi_M,\phi_B\,]$ 
is the action for the quantum field $\phi_M$
\begin{eqnarray}
 \label{action phiM sphere}
\hspace{-0cm}
S_{\textrm{P}^1,N,M}[ \,\phi_M,\phi_B\,] &=& \nonumber \\
 & & \hspace{-2.4cm}
\int_{\textrm{P}^1}  \left[\,\frac{1}{\pi}\,
  \partial_z  \phi_M\,
  \partial_{\bar{z}}\phi_M +\mu_\textrm{g} e^{2b_\textrm{g} \phi_B}
\Big(e^{2b_\textrm{g}(\,\phi_M+
 \phi_0\,)}-1\Big)
  - \frac{ 2\,(\,1-\hat\alpha\,)}{\pi}\,\phi_M \,  \partial_z
  \partial_{\bar{z}} \phi_B \, \right]  d ^2z~. \nonumber \\
 & &
\end{eqnarray}
\noindent Notice that, for $Q=1/b_\textrm{g}$, $e^{2b_\textrm{g}\phi}$ 
is a $(1,1)$ density at the classical level, 
which can be achieved by assigning to $e^{2b_\textrm{g}\phi_B}$ 
a classical $(1,1)$
nature and treating $\phi_0$ and $\phi_M$ as scalars.\\
The background action $S_{\textrm{P}^1,B}[\,\phi_B\,]$, the action 
$S_{\textrm{P}^1,N,M}[ \,\phi_M,\phi_B\,]$ for the quantum field $\phi_M$
and the term
\begin{equation}
 \label{(2-alpha) term}
 - \, \frac{ 2-\hat\alpha}{\pi} \,\int_{\textrm{P}^1} \phi_0 \,  
\partial_z \partial_{\bar{z}} \phi_B \, d ^2z
\end{equation}
are invariant 
under a  $SL\,(\,2,\mathbb{C}\,)$ transformation
\begin{equation}
 \label{SL(2C) transformation}
z \hspace{.2cm}\longrightarrow \hspace{.2cm}w= \frac{az+b}{cz+d} 
\hspace{1cm}\qquad 
\left( \begin{array}{ll} a & b \\
                         c & d \end{array} \right) \in SL (\,2,\mathbb{C}\,)
\end{equation}
when $Q=1/b_\textrm{g}$.\\
Thus, the transformation properties of
$S_{\textrm{P}^1,\,N}[ \,\phi\,]$ can be read from the second line
of (\ref{geometric action Sphere phiM}).\\
Taking into account that, being $\phi_0$ a scalar field, $c_0$
transforms like
\begin{equation}
 \label{c'_0}
c'_0=c_0- \sum_{n=1}^N \alpha_n \, \log \,|\, cz_n+d\,|^2
\end{equation}
one obtains
\begin{equation}
 \label{action sphere transformation}
S'_{\textrm{P}^1, \,N}[ \,\phi'\,] = S_{\textrm{P}^1, \,N}[\,\phi\,] + 
\sum_{n=1}^N \alpha_n\,\left(\,Q-\alpha_n\,\right)\, 
\log \,\left|\, \frac{d w}{d z} \, \right|_{z=z_n}^2.
\end{equation}
where 
$\log \left|\,d w/d z \, \right|_{z=z_n}^2=-\,2 \log |\,c z_n+d\,|^2$.\\
We remark 
that the geometric action (\ref{geometric action Sphere phiM}) has the
transformation property (\ref{action sphere transformation}) under
$SL\,(\,2,\mathbb{C}\,)$ only for
$Q=1/b_\textrm{g}$.\\
The crucial property of the geometric action is to provide, through
(\ref{action sphere transformation}) and the 
$SL\,(\,2,\mathbb{C}\,)$ invariant measure,
the quantum conformal dimensions
\begin{equation}
 \label{conformal dimensions sphere}
\Delta_{\alpha} = \alpha\,\left(\,Q-\alpha\,\right)=
\alpha\,\left(\,\frac{1}{b_\textrm{g}}-\alpha\,\right)
\end{equation}
for the Liouville vertex operators $e^{2\alpha\phi(z)}$.\\

\end{document}